\documentclass[universe,article,accept,moreauthors,pdftex]{Definitions/mdpi}

\firstpage{1}
\pubvolume{1}
\issuenum{1}
\articlenumber{0}
\pubyear{2024}
\copyrightyear{2024}
\datereceived{31 December 2023}
\daterevised{8 February 2024} 
\dateaccepted{12 February 2024}
\datepublished{ }
\hreflink{https://doi.org/} 

\usepackage{epsfig}
\usepackage{amssymb}
\usepackage[utf8]{inputenc}
\usepackage{graphicx}
\usepackage{dcolumn}
\usepackage{bm}
\newcommand{\be}{\begin{equation}}
\newcommand{\ee}{\end{equation}}
\newcommand{\bea}{\begin{eqnarray}}
\newcommand{\eea}{\end{eqnarray}}
\newcommand{\ba}{\begin{array}}
\newcommand{\ea}{\end{array}}
\newcommand{\bi}{\begin{itemize}}
\newcommand{\ei}{\end{itemize}}
\newcommand{\lan}{\langle}
\newcommand{\ran}{\rangle}

\renewcommand{\vec}[1]{\mbox{\boldmath $#1 \!\!$ \unboldmath}}

\Title{Induced Isotensor Interactions in Heavy-Ion Double-Charge-Exchange Reactions and the Role of Initial and Final State Interactions} 
\TitleCitation{Induced Isotensor Interactions in Heavy-Ion Double-Charge-Exchange Reactions and the Role of Initial and Final State Interactions}

\Author{Horst Lenske 
 $^{1,*}$$^{,\dagger}$\orcidA{0000-0002-9188-6032}, Jessica Bellone $^{2,\dagger}$, Maria Colonna $^{2,\dagger}$, Danilo Gambacurta $^{2,\dagger}$ and Jos\'{e}-Antonio Lay $^{3,4,\dagger}$}

\AuthorNames{Horst Lenske, Jessica Bellone, Maria Colonna and Danilo Gambacurta}

\AuthorCitation{Lenske, H.; Bellone, J.; Colonna, M.; Gambacurta, D.; Lay, J.-A.}

\address{%
$^{1}$ \quad {Institut f\"ur Theoretische Physik, Justus--Liebig--Universit\"at Giessen}, {D-35392 Giessen}, Germany\\

$^{2}$ \quad {Laboratori Nazionali del Sud, Istituto Nazionale di Fisica Nucleare}, I-95123 Catania, Italy;\newline {bellone@lns.infn.it (J.B.); colonna@lns.infn.it (M.C.); gambacurta@lns.infn.it (D.G.)}\\

$^{3}$ \quad {Departamento de FAMN, Facultad de Física, Universidad de Sevilla, Apartado 1065, E-41080 Sevilla}, Spain\\
$^{4}$ \quad {Instituto Interuniversitario Carlos I de F\'isica Te\'orica y Computacional (iC1), Apdo.~1065, E-41080 Sevilla}, Spain; lay@us.es\\
$^{\dagger}$ \quad The NUMEN Collaboration, LNS Catania, I-95123 Catania, Italy.}

\corres{Correspondence: horst.lenske@physik.uni-giessen.de; Tel.: +49 641 9933361}

\date{\today}

\abstract{The role of initial state (ISI) and final state (FSI) ion--ion interactions in heavy-ion double-charge-exchange (DCE) reactions $A(Z,N)\to A(Z\pm 2,N\mp 2)$ are studied for double single-charge-exchange (DSCE) reactions given by sequential actions of the isovector nucleon--nucleon (NN) T-matrix. In momentum representation, the second-order DSCE reaction amplitude is shown to be given in factorized form by projectile and target nuclear matrix elements and a reaction kernel containing ISI and FSI. Expanding the intermediate propagator in a Taylor series with respect to auxiliary energy allows us to perform the summation in the leading-order term over intermediate nuclear states in closure approximation. 
The nuclear matrix element attains a form given by the products of two-body interactions directly exciting the $n^2p^{-2}$ and $p^2n^{-2}$ DCE transitions in the projectile and the target nucleus, respectively. 
A surprising result is that the intermediate propagation induces correlations between the transition vertices, showing that DSCE reactions are a two-nucleon process that resembles a system of interacting spin--isospin dipoles. Transformation of the DSCE NN T-matrix interactions from the reaction theoretical t-channel form to the s-channel operator structure required for spectroscopic purposes is elaborated in detail, showing that, in general, a rich spectrum of spin scalar, spin vector and higher-rank spin tensor multipole transitions will contribute to a DSCE reaction. Similarities (and differences) to two-neutrino double-beta decay (DBD) are discussed. ISI/FSI distortion and absorption effects are illustrated in black sphere approximation and in an illustrative application to data.}


\begin{document}

\tableofcontents 

\section{Introduction}  
A new era of nuclear spectroscopy with heavy-ion beams has begun by the systematic use of heavy-ion single-charge-exchange (SCE) and, very recently, double-charge-exchange (DCE) reactions. While SCE studies reach back to the last century, DCE investigations are a rather new application of heavy-ion beams for spectroscopic studies. As reviewed in recent articles~\cite{Lenske:2019cex,Cappuzzello:2022ton}, over the last decade, DCE research has evolved rapidly, both on the experimental and the theoretical sides. Yet, compared to the status of SCE reaction physics, heavy-ion DCE research is still in its infancy. Going back in history, one finds that early studies of heavy-ion-induced DCE reactions were not pursued further after the first measurements at Los Alamos~\cite{Drake:1980}, GANIL at Caen~\cite{Naulin:1982} and NSCL at Michigan State University~\cite{Blomgren:1995cux} led to
poor yields and conflicting results. The large differences in the reported cross sections
have never been explored systematically: neither experimentally nor theoretically. Not in the least, the lack of an
adequate theoretical framework for nuclear DCE reactions inhibited a clear interpretation of the data, leaving the question
unsolved as to what degree the final DCE channels were indeed produced by sequential multi-nucleon transfer processes, which were the theoretically favored explanations for such reactions at the time of the early DCE data~\cite{Bes:1983tty,Dasso:1985zvv,Dasso:1986zza,vonOertzen:1995uba}, leaving open the question of whether there are other competing reaction mechanisms.

From our recent theoretical investigations summarized in~\cite{Lenske:2018jav,Lenske:2019cex,Cappuzzello:2022ton}, a more complete and complex picture of heavy-ion DCE reactions emerged.
In general, heavy-ion DCE reactions are determined by three competing reaction mechanisms with different dynamical origins and spectroscopic content. Multi-step transfer reactions are always involved but can be suppressed to a negligible level by an appropriate choice of projectile and target nuclei and by performing experiments at incident energies sufficiently high above the Coulomb barrier. Transfer processes are maintained by the nuclear mean-field. As such, they are probing, in the first place, single-particle and nucleon-pair dynamics without directly accessing the isovector response of nuclei and the intrinsic isospin properties of nucleons as of interest for spectroscopic work and nuclear beta decay.

The nucleonic degrees of freedom beyond the mean-field are probed only in collisional NN interactions involving explicitly isovector NN interactions. In fact, nuclear-charge-exchange reactions will always contain collisional NN contributions. As in heavy-ion SCE reactions, the conditions of DCE reactions may be chosen as to enhance the collisional components. However, in practice, that will require researchers to always study the transfer route and the collisional route to the final DCE channel in order to keep as much as possible full control over the reaction. Of high importance for reliable reaction calculations are high-quality optical potentials, which always requires in parallel measurements of elastic scattering cross sections, 
at least for the incident channel configuration. The multi-method approach pursued by the NUMEN collaboration is based on such a far-reaching scheme and experimentally and theoretically investigates elastic scattering and the various intermediate transfer and collisional SCE reactions that possibly contribute as intermediate states to a DCE reaction.

\begin{figure}[H]

\includegraphics[width = 13.5cm]{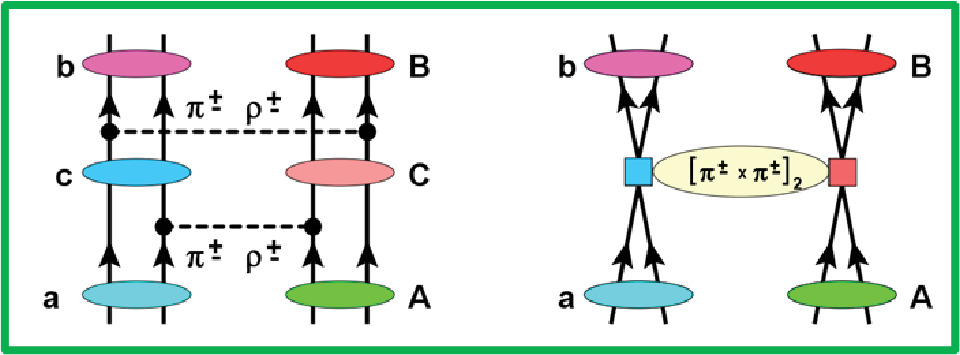}
\caption{Schematic representation of  the collisional processes contributing to a DCE reaction $A(Z,N)\to B(Z\pm 2,N\mp 2)$. The DSCE reaction scenario of second-order in the isovector NN T-matrix (left) competes with the direct MDCE mechanism proceeding by an isotensor interaction induced by off-shell pion--nucleon DCE scattering. }
\label{fig:DCE}

\end{figure}

As depicted in Figure \ref{fig:DCE}, in a heavy-ion DCE reaction, the isospin structure of hadrons and the isospin response of the nuclear medium can be probed in two ways: either by sequential action of the NN isovector interactions as a double single-charge-exchange (DSCE) reaction or in a virtual meson--nucleon double-charge-exchange reaction as a Majorana DCE (MDCE) reaction~\cite{Lenske:2018jav,Lenske:2021jnr}. The MDCE mechanism is dominated by pion DCE. Utilizing the meson-exchange picture of nuclear interactions, the two collisional DCE mechanisms have in common that charge conversion proceeds by the exchange of virtual charged mesons between the reacting nuclei. An MDCE reaction is of first-order in an isotensor interaction dynamically generated by a pair of virtual $\pi^\pm,\pi^\mp$ reactions. The peculiarities of MDCE reactions will be discussed in a separate paper. In this work, DSCE reactions will be considered. They are second-order reactions in which the final DCE channel is populated in a two-step manner by acting twice with the NN isovector T-matrix.

\textls[-15]{An appealing aspect of DCE physics is the possibility to probe the same nuclear configurations as encountered in double-beta-decay (DBD). That topic is gaining new and wide attention: see, e.g.,~\cite{Ejiri:2022ujl}. A comprehensive introduction and overview on DBD physics was given by Tomoda~\cite{Tomoda:1990rs}. The present status of DBD research was reviewed by Ejiri et al.~\cite{Ejiri:2019ezh} and Agostini et al.~\cite{Agostini:2022zub}. Cutting the meson lines vertically in the DSCE diagram in Figure~\ref{fig:DCE}, a striking diagrammatical similarity of DSCE reactions to two-neutrino ($2\nu 2\beta$) DBD is discovered. However, dynamically, the two charge-converting processes are fundamentally different although they proceed by the same kind of spin--isospin operators. The origins and strengths of the underlying dynamics and the intrinsic structure of the interaction vertices are quite different, and the nuclear and weak coupling constants differ by orders of magnitude. But it is worthwhile to recall the established, fruitful connection between nuclear  SCE reactions and single-neutrino beta decay (SBD)~\cite{Akimune:2020utu,Frekers:2016haa,Frekers:2017kta,Frekers:2018}. Light-ion SCE reactions were, in fact, used to indirectly collect information on DBD nuclear matrix elements~\cite{Puppe:2011zz,Thies:2012xg,Thies:2012zz,Ejiri:2019ezh,Ejiri:2020xmm}. In combination, nuclear SCE and DCE reactions are the only known processes that allow the simultaneous study of the rank-1 isovector and the rank-2 isotensor responses of nuclei in laboratory experiments under reproducible conditions. It is worthwhile to mention that both kinds of collisional DCE reactions proceed by effective four-body isotensor interactions acting as two-body interactions in the participating nuclei. Heavy-ion DCE reactions are unique for providing for the first time the opportunity to investigate such high-rank operators.}

Although in about the past two decades a lot of experience has been collected  on multi-particle-hole dynamics in low-lying nuclear excitations---see e.g.,~\cite{Litvinova:2013lxa,Fukui:2017lrt,Lenske:2019ubp,Soderstrom:2020iaz} for experiments and theories on double-photon emission---only a little work has been spent on systematic investigations of nuclear DCE spectroscopy. Possible relations between DBD and double-photon processes were investigated quite recently by Jokiniemi and Menendez~\cite{Jokiniemi:2023bes}. For such comparative studies, heavy-ion DCE reactions offer new opportunities by giving access to another class of double-excitation processes driven by nuclear interactions.

In a previous article~\cite{Lenske:2021jnr}, we have already discussed the theoretical methods required to convert the second-order DSCE reaction results, for which a description in a t-channel formalism is the natural approach, to second-order nuclear matrix elements, which are s-channel objects. As an independent additional issue, we present in this article an alternative approach by performing that transformation directly on the operator level. That method has the advantage of obtaining deeper insight into the DCE dynamics and the interplay of reaction and structure physics. As an important result, we derive the effective isotensor interactions that are 
generated in a DSCE reaction dynamically in a cooperative manner by the colliding ions.

The mentioned aspects will be discussed here for the DSCE reaction mechanism, but the ISI/FSI complex and the obtained results are of general relevance for all nuclear reactions of second-order distorted wave character. Similar features are also encountered in reactions proceeding by the MDCE mechanism; however, this is deserving of separate consideration because of the vastly different reaction dynamics. The MDCE scenario will be addressed separately in a follow-up paper.

The central goal of this paper is the “proof of principle” of how to extract from a heavy-ion DCE reaction the spectroscopic information under the conditions of strong ISI and FSI. Anticipating the result, such a separation is indeed possible in momentum representation where the second-order DSCE reaction amplitude is obtained as a folding integral of nuclear transition form factors and an ISI/FSI distortion kernel. Further progress is made by the closure approximation, for which an estimate of the neglected terms by sum rules is derived. Investigations of the intermediate propagator reveal that short-range correlations are induced between the two pairs of SCE vertices in the interacting nuclei, thus indicating that DSCE transitions are in fact two-nucleon processes. An important step for the proper definition of nuclear matrix elements is the transformation of the NN isovector interactions from the  
t-channel formulation used in reaction theory to the s-channel formalism appropriate for nuclear structure investigations. As a relatively easy to handle case, the formalism is used to evaluate the DSCE reaction amplitude in the Black Sphere (BS) strong absorption limit, which is appropriate for heavy-ion reactions. 

In the following, we address two key questions related to DSCE theory: 
\begin{itemize}
  \item What is the role of initial state (ISI) and final state (FSI) ion--ion interactions in spectroscopic studies?
  \item How can we extract nuclear matrix elements from heavy-ion double-charge-exchange reactions?
\end{itemize}
Both issues are of vital importance for the aim of using heavy-ion DCE reactions as a spectroscopic tool, not the least as surrogate reactions for double-beta decay. Both questions are investigated in the context of double single-charge-exchange reactions 
treated as second-order processes of the NN isovector T-matrix.

In Section \ref{sec:DSCE}, second-order DW reaction theory is recapitulated, and the DSCE reaction amplitude is presented in standard second-order form. The main result of Section \ref{sec:DSCE}, however, is the introduction of the closure approximation for the DSCE amplitude. In Section \ref{sec:Kernel}, an alternative, new formulation is presented for the DSCE reaction amplitude utilizing the momentum representation. The most important aspect is the separation of the nuclear structure and reaction dynamics. Section \ref{sec:DSCE_NME} is devoted to the DSCE-NME. The t-channel operator structure of the product of T-matrices is recast into s-channel two-body operators acting in the projectile and target nucleus. We reconsider our previous formulation of the problem in~\cite{Lenske:2021jnr} now on the level of operators rather than on the level of matrix elements. The various aspects of the formalism are illustrated in  Section \ref{sec:IllusDSCE}. The discussed issues are summarized, conclusions are drawn and an outlook to future work is given in Section \ref{sec:Sum}. The multipole structure of the effective induced isotensor interaction is presented in a number of the appendices.

\section{Reaction Theory of Double Single-Charge-Exchange Reactions}\label{sec:DSCE}

\subsection{The DSCE Reaction Amplitude and the Intermediate Propagator}

DSCE reactions leading from the incident channel $\alpha= a(Z_a,N_a)+A(Z_A,N_A)$ to the exit channel $\beta=b(Z_a\pm 2,N_a\mp 2)+B(Z_A\mp 2,N_A\pm 2)$ are a sequence of two single-charge-exchange events, each of them mediated by the two-body NN--isovector interaction $\mathcal{T}_{NN}$. The many-body wave function $\Psi^{(+)}_{aA}$ with asymptotically outgoing spherical waves is expanded  into the complete set of asymptotic eigenstates of the $a$ and the $A$ systems, respectively. The reaction amplitude is then written down readily as a quantum mechanical second-order matrix element
\cite{Bellone:2020lal}:
\be\label{eq:MDSCE}
\mathcal{M}^{(2)}_{\alpha\beta}(\mathbf{k}_\alpha,\mathbf{k}_{\beta})=
\lan \chi^{(-)}_\beta, bB|\mathcal{T}_{NN}\mathcal{G}^{(+)}_{aA}(\omega_\alpha)\mathcal{T}_{NN}|aA,\chi^{(+)}_{\alpha} \ran  .
\ee

The DSCE differential cross section for unpolarized ions is given as
\be\label{eq:dsigma}
d\sigma^{(2)}_{\alpha\beta}=\frac{m_\alpha m_\beta}{(2\pi\hbar^2)^2}\frac{k_\beta}{k_\alpha}\frac{1}{(2J_a+1)(2J_A+1)}
\sum_{M_a,M_A\in \alpha;M_b,M_B\in \beta}{\left|M^{(2)}_{\alpha\beta}(\mathbf{k}_\alpha,\mathbf{k}_\beta)\right|^2}d\Omega,
\ee
averaged over the initial nuclear spin states ($J_{a,A},M_{a,A}$) and summed over the final nuclear spin states  ($J_{b,B},M_{b,B}$), respectively. Reduced masses in the incident and exit channel, respectively, are denoted by $m_{\alpha,\beta}$. The variables $\mathbf{k}_\alpha$ and $\mathbf{k}_\beta$  are the (invariant) momenta in the incident and exit channels, respectively.

Initial (ISI) and final state (FSI) interactions are taken into account by optical potentials. The distorted waves $\chi^{(\pm)}_{\alpha,\beta}$ depend on the center-of-mass (c.m.) momenta $\mathbf{k}_{\alpha,\beta}$ and obey outgoing and incoming spherical wave boundary conditions, respectively.  Denoting the projectile and target nucleus by $a$ and $A$, respectively, the energy available in the center-of-mass rest frame $s_{\alpha}=(T_{lab}+M_a+M_A)^2-T_{lab}(T_{lab}+2M_a)$. In the following, we use $\omega_\alpha=\sqrt{s_{\alpha}}$. The perturbative approach of Equation  \eqref{eq:MDSCE} is justified by the weak direct coupling of the DSCE channels to other reaction channels.

The many-body Green's function of the combined projectile--target system is denoted by $\mathcal{G}^{(+)}_{aA}$. In operator form, it is given by the resolvent of the projectile plus target many-body Schr\``{o}dinger equation. Since DCE reactions are peripheral, grazing collisions, the $a+A$ system remains in each reaction step separable into projectile and target nuclei, allowing us to express the total Hamiltonian by the sum of the nuclear Hamiltonians $H_{a}$ and $H_{A}$ plus the Hamiltonian $H_{aA}=H_{rel}+V_{aA}$. $H_{rel}=T_{cm} +U_{aA}$ accounts for the kinetic energy of relative motion ($T_{cm}$) and elastic ion--ion interactions ($U_{aA}$), which for the assumed reaction scenario are well-approximated by an optical model Hamiltonian $H_{opt}$. The residual projectile--target interactions are approximated by the nucleon--nucleon (NN) T-matrix, $V_{aA}\approx \mathcal{T}_{NN}$; thus, the full Lippmann--Schwinger scattering series of NN-interactions are included. 
In a fully microscopic approach as in~\cite{Bellone:2020lal}, $U_{aA}$ is calculated by a folding approach by using $\mathcal{T}_{NN}$ and nuclear ground state densities, which are derived self-consistently from energy density functionals by the Hartree--Fock--Bogolyubov (HFB) theory~\cite{Lenske:2018jav}.

\begin{figure}[H]

\includegraphics[width = 10cm]{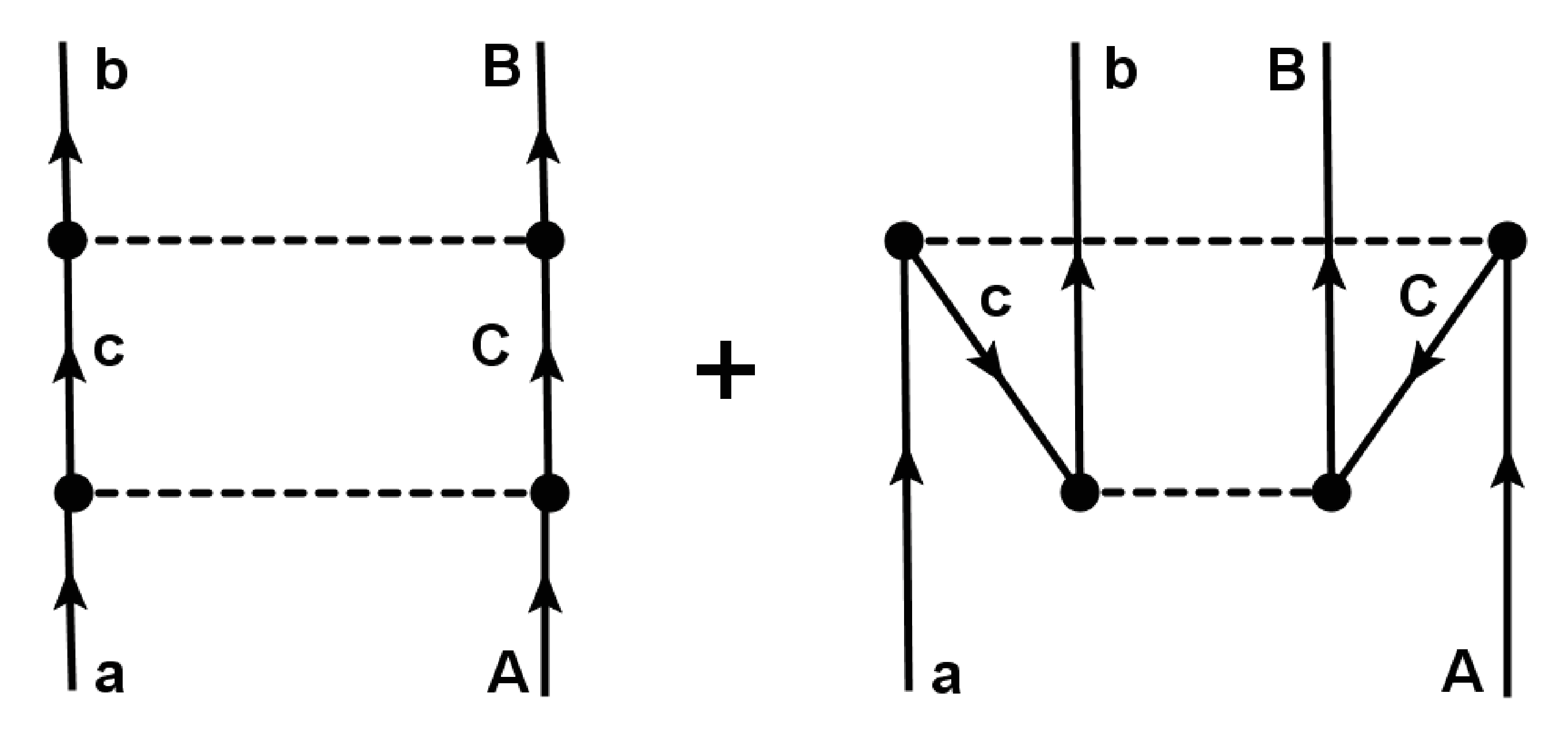}
\caption{The DSCE ladder and the so-called Z graph entering into the retarded DSCE propagator. }
\label{fig:DSCE_prop}

\end{figure}

As shown in Figure \ref{fig:DSCE_prop}, the proper treatment of time ordering is taken care of by using the retarded propagator:
\be
\mathcal{G}^{(+)}_{aA}(\omega_\alpha)=
\frac{1}{\omega_\alpha-H_a-H_A-H_{opt}+i\eta}+\frac{1}{\omega_\alpha+H_a+H_A+H_{opt}+i\eta}.
\ee
where the first and second terms correspond to the ladder graph and the so-called Z graph, respectively; $\eta \to 0+$ denotes an infinitesimal approaching zero from positive values, which 
ensures outgoing spherical wave boundary conditions.

The basis of states $c \in \{a\}$ and $C\in \{A\}$ are solutions of the nuclear Hamiltonians $(H_a-E_c)|c\ran=0$ and $(H_A-E_C)|C\ran=0$, respectively. Hence, we obtain the representation:
\be\label{eq:GaA}
\mathcal{G}^{(+)}_{aA}(\omega_\alpha)=\sum_{\gamma=cC}|cC\ran G^{(+)}_{\alpha\gamma}(\omega_\alpha)\lan cC| .
\ee

The channel propagators
\be\label{eq:GcC}
G^{(+)}_{\alpha\gamma}=\frac{1}{\omega_\alpha-M_\gamma-H_{opt}+i\eta}+\frac{1}{\omega_\alpha+M_\gamma+H_{opt}+i\eta}
\ee
describe the evolution of the system in a given intermediate partition $|\gamma\ran=|cC\ran$.  The energy $M_\gamma=M^*_c+M^*_C$ is given by the nuclear masses $M^*_{c,C}=M_{c,C}+\varepsilon_{c,C}$, which include the excitation energies $\varepsilon_{c,C}$.

$G^{(+)}_{\alpha\gamma}$ is expanded into the bi-orthogonal set
$\{\chi^{(+)}_\gamma,\widetilde{\chi}^{(+)*}_\gamma \}$ of optical model distorted waves (DWs) that are the solution of the wave equation defined by $H_{opt}$. Here, $\widetilde{\chi}^{(+)}_\gamma$ is the dual distorted wave obeying  $\lan\widetilde{\chi}^{(+)}_\gamma|\chi^{(+)}_{\gamma'}\ran=(2\pi)^3\delta(\mathbf{k}_\gamma-\mathbf{k}'_\gamma)\delta_{\gamma\gamma'}$
\cite{Lenske:2018jav,Bellone:2020lal,Lenske:2021bpk,Lenske:2021jnr}. This leads to the reduced propagators
\be \label{eq:redProp}
g^{(+)}_{\alpha\gamma}(k_\gamma)=\frac{1}{\omega_\alpha-M_\gamma-T_\gamma(k_\gamma)+i\eta}+
\frac{1}{\omega_\alpha+M_\gamma+T_\gamma(k_\gamma)+i\eta}
\ee
with the kinetic energy $T_\gamma(k_\gamma)=\sqrt{k^2_\gamma+M^{*2}_c}+\sqrt{k^2_\gamma+M^{*2}_C}-M_\gamma\sim \frac{k^2_\gamma}{2m_\gamma}$ at the off-shell momentum $k_\gamma$; $1/m_\gamma=1/M^*_c+1/M^*_C$ is the reduced mass.
The  DSCE reaction amplitude is rewritten as
\be\label{eq:MDSCE_M1M1}
\mathcal{M}^{(2)}_{\alpha\beta}(\mathbf{k}_\alpha,\mathbf{k}_\beta)=\sum_{\gamma=\{c,C\}}
\int \frac{d^3k_\gamma}{(2\pi)^3}M^{(1)}_{\gamma\beta}(\mathbf{k}_\gamma,\mathbf{k}_\beta)
g^{(+)}_{\alpha\gamma}(k_\gamma)
\widetilde{M}^{(1)}_{\alpha\gamma}(\mathbf{k}_\alpha,\mathbf{k}_\gamma),
\ee

The non-Hermitian nature of $H_{opt}$ has led to two distinct half off-shell first-order DW amplitudes~\cite{Lenske:2021bpk,Lenske:2019cex,Bellone:2020lal}:
\bea
\widetilde{M}^{(1)}_{\alpha\gamma}(\mathbf{k}_\alpha,\mathbf{p}_\gamma)=
\lan \widetilde{\chi}^{(+)}_\gamma,cC|T_{NN}|aA,\chi^{(+)}_\alpha\ran\\
M^{(1)}_{\gamma\beta}(\mathbf{p}_\gamma,\mathbf{k}_\beta)=
\lan \chi^{(-)}_\beta,bB|T_{NN}|cC,\chi^{(+)}_\gamma\ran .
\eea

\subsection{DSCE Reaction Amplitude in Momentum Representation}

The structure of the DSCE transition form factor becomes especially transparent in momentum space.
In momentum representation, the anti-symmetrized, energy-dependent isovector T-matrix, acting between a nucleon in $i \in \{a\}$ and a nucleon in $j \in \{A\}$ is given by the 3D Fourier transform~\cite{Lenske:2018jav,Lenske:2021jnr}
\be \label{eq:TNN_FB}
\mathcal{T}_{NN}(i,j)=\int \frac{d^3p}{(2\pi)^3}e^{i\mathbf{p}\cdot \mathbf{x}_{ij}}\mathcal{T}_{NN}(\mathbf{p}).
\ee

In the ion--ion rest frame, the relative distance between nucleons in the projectile and target nucleus is  $\mathbf{x}_{ik}=\mathbf{r}_k-\mathbf{r}_i+\mathbf{r}_{\lambda}$ and 
is given by the intrinsic nuclear coordinates $\mathbf{r}_{i,k}$ and the ion--ion relative coordinate $\mathbf{r}_\lambda$, $\lambda=\alpha,\beta$---the latter describing the distance between the centers-of-mass of the projectile-like and the target-like nuclei.

The isovector ($T=1$) part of the NN T-matrix consists of central  rank-0 spin scalar and rank-1 spin vector and 
non-central rank-2 spin tensor interactions:
\be\label{eq:TNN_momentum}
\mathcal{T}_{NN}(\mathbf{p}|ik)=
\sum_{S=0,1;T=1}
\left(V_{ST}(p^2)\left[\bm{\sigma}_i\cdot\bm{\sigma}_k\right]^S+\delta_{S1}V_{Tn}(p^2)Y_2(\hat{\mathbf{p}})\cdot \left[\bm{\sigma}_i\otimes\bm{\sigma}_k\right]_2
\right)\bm{\tau}_i\cdot\bm{\tau}_k.
\ee
where only the isospin ladder parts $\tau^\pm_{i,k}$ are relevant for the SCE process.
The (complex) form factors $V_{ST}$ and $V_{Tn}$ are functions of the momentum transfer $\mathbf{p}$ and the energies of the interacting nucleons 
and depend on the density of the surrounding medium.
$Y_{2M}(\hat{\mathbf{p}})$ is a rank-2 spherical harmonic that is contracted with the previously mentioned rank-2 spin tensor. 
The spin scalar parts excite Fermi-type (F) transitions and the spin vector and spin tensor components induce Gamow--Teller-type (G) modes mutually in the projectile and target.

In momentum representation, SCE transitions are described by matrix elements of the scattering operators:
\bea\label{eq:SNN}
\mathcal{S}_{NN}(\mathbf{p}|ik)=e^{i\mathbf{p}(\mathbf{r}_i-\mathbf{r}_k)}T_{NN}(\mathbf{p}|ik)\tau_\pm(i)\tau_{\mp}(j).
\eea

The plane-wave factor accounts for the spatial multipole structure, resulting in the Bessel--Fourier transforms of the one-body transition densities constructed from the $np^{-1}$ r $pn^{-1}$ particle-hole configuration amplitudes and the corresponding single-particle wave functions~\cite{Lenske:2018jav}.
Odd indices $i=1,3\ldots$ and even indices $k=2,4\ldots$ will be used to denote states in the $a$- and the $A$-system, respectively.

\subsection{Distortion and Absorption}
The DSCE amplitude of Equation \eqref{eq:MDSCE} is the standard form for a DW matrix element in second-order perturbation theory in $\mathcal{T}_{NN}$. In momentum representation, the DSCE reaction amplitude is retrieved, however, in a rather different form:
\be\label{eq:M2_momentum}
\mathcal{M}^{(2)}_{\alpha\beta}(\mathbf{k}_\alpha,\mathbf{k}_\beta)=
\int d^3p_1\int d^3p_2\int \frac{d^3k_\gamma}{(2\pi)^3}
\widehat{K}_{\alpha\beta}(\mathbf{p}_1,\mathbf{p}_2|\mathbf{k}_\gamma) \mathcal{N}_{\alpha\beta}(\mathbf{p}_1,\mathbf{p}_2) .
\ee

Remarkably, the spectroscopic and reaction contribution appear in factorized form. Nuclear spectroscopy is contained in the DSCE nuclear matrix element given by the form factor:
\be\label{eq:Nab}
\mathcal{N}_{\alpha\beta}(\mathbf{p}_1,\mathbf{p}_2)=
\sum_{cC}\lan bB|\mathcal{S}_{NN}(\mathbf{p}_2|(34))|cC\ran g^{(+)}_{\alpha\gamma}(k_\gamma)
                 \lan cC|\mathcal{S}_{NN}(\mathbf{p}_1 |(12))|aA\ran .
\ee

The NME resembles by its second-order structure  
formally the NME of $2\nu$-DBD. However, while in two-neutrino DBD, only intermediate $1^+$ Gamow--Teller states are of importance; in a DCE reaction, a much broader spectrum of intermediate states of other multipolarities will be excited, as was pointed out already in~\cite{Bellone:2020lal}.

Initial state (ISI) and final state (FSI) interactions are now completely contained in the reaction kernel:
\be
\widehat{K}_{\alpha\beta}(\mathbf{p}_1,\mathbf{p}_2|\mathbf{k}_\gamma)=
\widetilde{D}_{\alpha\gamma}(\mathbf{p}_1)D_{\gamma\beta}(\mathbf{p}_2)
\ee
which is determined by the
distortion coefficients~\cite{Lenske:2018jav}:
\bea\label{eq:DistortionCoeff}
\widetilde{D}_{\alpha\gamma}(\mathbf{p}_1)=\frac{1}{(2\pi)^3}\lan\widetilde{\chi}^{(+)}_\gamma|
e^{i\mathbf{p}_1\cdot \mathbf{r}_\alpha}|\chi^{(+)}_\alpha\ran\quad ; \quad
D_{\gamma\beta}(\mathbf{p}_2)=\frac{1}{(2\pi)^3}\lan\chi^{(-)}_\beta |
e^{i\mathbf{p}_2\cdot \mathbf{r}_\beta}|\chi^{(+)}_\gamma\ran .
\eea

\subsection{Closure Approximation}
If in Equation \eqref{eq:Nab}, the reduced channel propagator $g^{(+)}_{\alpha\gamma}(k_\gamma)$ is independent of the energies of the intermediate states, it can be extracted, which would allow us to perform the summations over an intermediate state closure.
For removing the channel dependence from the energy denominators $g^{(+)}_{\alpha\gamma}$, we add and subtract a state-independent auxiliary energy $\omega_\gamma$ and perform a Taylor-series expansion in
$\xi=1-\frac{M_\gamma}{\omega_\gamma}$. In leading-order, the propagator has become independent of the intermediate states, which allows us to perform the summations over the channel states $c,C$. As a result, we obtain the DSCE NME in a closure approximation:
\be
\mathcal{F}_{\alpha\beta}(\mathbf{p}_1,\mathbf{p}_2)=\sum_{(13)\in \{a\}}\sum_{(24)\in \{A\}}
\lan bB| \mathcal{R}_{NN}(\mathbf{p}_1,\mathbf{p}_2|13,24)|aA\ran .
\ee

The initial and final nuclear states are connected directly by the effective DSCE rank-2 isotensor interaction:
\be\label{eq:RNN}
\mathcal{R}_{NN}(\mathbf{p}_1,\mathbf{p}_2|13,24)=
\mathcal{S}_{NN}(\mathbf{p}_2|34)\mathcal{S}_{NN}(\mathbf{p}_1|12).
\ee

The DSCE transition operator $\mathcal{R}_{NN}$ is a four-body operator composed of the products of two-body operators: one acting in the projectile and the other in the target nucleus.

Attaching the modified, state-independent propagator to the reaction kernel and also including the $k_\gamma$ integration, we obtain
\be \label{eq:Kernel}
\mathcal{K}_{\alpha\beta}(\mathbf{p}_1,\mathbf{p}_2)=\int \frac{d^3k_\gamma}{(2\pi)^3}
D_{\gamma\beta}(\mathbf{p}_2)g^{(+)}_{\alpha\gamma}(k_\gamma).
\widetilde{D}_{\alpha\gamma}(\mathbf{p}_1).
\ee

The closure approach renders Equation  \eqref{eq:M2_momentum} into an intriguing simple form:
\be\label{eq:M2_closure}
\mathcal{M}^{(2)}_{\alpha\beta}(\mathbf{k}_\alpha,\mathbf{k}_\beta)=
\int d^3p_1\int d^3p_2\mathcal{K}_{\alpha\beta}(\mathbf{p}_1,\mathbf{p}_2)
\mathcal{B}_{\alpha\beta}(\mathbf{p}_1,\mathbf{p}_2)  +Res(\xi).
\ee

Terms of first- and higher-order in $\xi$ are contained in the residual term $Res(\xi)$, which is discussed further in Appendix \ref{app:Residual}.

\section{The DSCE Reaction Kernel and Initial and Final State Interactions}\label{sec:Kernel}

\subsection{The Reaction Kernel in Plane-Wave Approximation}

An instructive limiting case is to neglect all elastic interactions, leading to the plane-wave approximation (PWA). In the PWA, the distortion coefficients, Equation  \eqref{eq:DistortionCoeff}, achieve a particular simple form: $\widetilde{D}_{\alpha\gamma}=\delta(\mathbf{q}_{\alpha\gamma}+\mathbf{p}_1)$ and
$D_{\gamma\beta}=\delta(\mathbf{\mathbf{q}}_{\gamma\beta}+\mathbf{p}_2)$, respectively, where $\mathbf{q}_{\alpha\gamma}=\mathbf{k}_\alpha-\mathbf{k}_\gamma$ and $\mathbf{q}_{\gamma\beta}=\mathbf{k}_\gamma-\mathbf{k}_\beta$ are the half off-shell momentum transfers in the first and second SCE interactions. The total on-shell momentum transfer is $\mathbf{q}_{\alpha\beta}=\mathbf{q}_{\alpha\gamma}+\mathbf{q}_{\gamma\beta}$.

Changing the momentum coordinates to $\mathbf{p}_{1,2}\mapsto \mathbf{P}\pm \mathbf{q}/2$ and introducing the mean on-shell channel momentum $\mathbf{P}_{\alpha\beta}=(\mathbf{k}_\alpha+\mathbf{k}_\beta)/2$, the product of PWA distortion factors and the $k_\gamma$-integral are easily evaluated, resulting in the PWA reaction kernel:
\be\label{eq:PW_Kernel}
\mathcal{K}^{(0)}_{\alpha\beta}(\mathbf{P},\mathbf{q})=
\frac{1}{8}\delta(\mathbf{P}+\frac{1}{2}\mathbf{q}_{\alpha\beta})
g^{(+)}_{\alpha\gamma}(|\mathbf{P}_{\alpha\beta}-\mathbf{q}/2|).
\ee

Both the delta-distribution and the functional structure of the propagator impose constraints on the momenta $\mathbf{P}$ and $\mathbf{q}$.\endnote{The factor $\frac{1}{8}$ is due to the extraction of the factor 2  from the delta-distribution, which is 
defined in 3-D momentum space.}
The above result also indicates that $\mathbf{P}=\frac{1}{2}(\mathbf{p}_1+\mathbf{p}_2)$ is closely related to the physical momentum transfer $\mathbf{q}_{\alpha\beta}$, which is exact in the PW limit but approximate with ISI/FSI.

\subsection{Distortion and Absorption}\label{ssec:DisAbs}

In order to gain insight into the effects introduced by ISI/FSI interactions, the product of distorted waves is factorized into a plane-wave part and a reduced, generally 
complex-valued, amplitude:
\bea
\widetilde{\chi}^{(+)\dag}_\gamma(\mathbf{r}_\alpha) \chi^{(+)}_\alpha(\mathbf{r}_\alpha)&\mapsto&
e^{i(\mathbf{k}_\alpha-\mathbf{k}_\gamma)\cdot \mathbf{r}_\alpha}\widetilde{\eta}_{\alpha\gamma}(\mathbf{r}_\alpha)\\
\chi^{(-)\dag}_\beta(\mathbf{r}_\beta) \chi^{(+)}_\gamma(\mathbf{r}_\beta)&\mapsto&
e^{i(\mathbf{k}_\gamma-\mathbf{k}_\beta)\cdot \mathbf{r}_\beta}\eta_{\gamma\beta}(\mathbf{r}_\beta).
\eea
For strongly absorbing systems like colliding heavy ions, the amplitudes $\eta$ and $\widetilde{\eta}$  are strongly suppressed in the overlap region, as is easily verified in Wentzel--Kramers--Brillouin (WKB) approximation or eikonal theory.
That region, centered at the origin, corresponds to an avoided volume in the sense that the spatial density distributions of the incoming and outgoing distorted waves vanishes. This kind of localized \emph{absorption of probability flux}
implies that the (quantum mechanical) survival probability of the initial system in that region approaches zero. As a consequence, reactions will be restricted to regions outside of the ion--ion overlap volume; these regions are characterized by 
strong absorption of probability flux due to coupling to other reaction channels.

The shape and extent 
of that void is imprinted in the reduced amplitude. Describing the shape and size of the wave-function voids by the absorption form factors $|h_{ij}(\mathbf{r})|$, we obtain  $\eta_{ij}(\mathbf{r})=1-h_{ij}(\mathbf{r})$.  For heavy-ion reactions, $|h_{ij}(\mathbf{r})|$ is unity up to the absorption radius $r=R_{abs}$ and decreases rapidly at larger radii~\cite{Lenske:2018jav}: in many aspects resembling a Heaviside distribution $h_{ij}( \mathbf{r})\sim \Theta(R_{abs}-r)$. In Ref.~\cite{Lenske:2018jav}, it was shown that $R_{abs}$ is directly related to the total ion--ion reaction cross section, and that it varies 
with the energy and mass numbers of the interacting nuclei.

In quasi-elastic reactions as considered here, the amplitudes $h_{ij}$ vary slowly with the energy and nuclear masses.  Thus, their properties are already fixed by the entrance channel, and we may use channel-independent, universal form factors: $h_{ij}(\mathbf{r})\equiv H_S(\mathbf{r})$ and $\widetilde{h}_{ij}\equiv \widetilde{H}_S(\mathbf{r}) $. Thus, we obtain
\bea
\widetilde{D}_{\alpha\gamma}(\mathbf{p}_1)&=&\delta(\mathbf{p}_1+\mathbf{q}_{\alpha\gamma})-
\widetilde{H}_{S}(\mathbf{p}_1+\mathbf{q}_{\alpha\gamma})\\
D_{\gamma\beta}(\mathbf{p}_2)&=&\delta(\mathbf{p}_2+\mathbf{q}_{\gamma\beta})-
H_{S}(\mathbf{p}_2+\mathbf{q}_{\gamma\beta}).
\eea

Hence, with ISI/FSI interactions, the distortion factors are given by subtracting the interactions occurring in the avoided volume from the maximal possible interaction probability of unity  described by the PWA delta-distribution.

The total DSCE reaction kernel is given by
\be
\mathcal{K}_{\alpha\beta}(\mathbf{p}_1,\mathbf{p}_2)=\mathcal{K}^{(0)}_{\alpha\beta}(\mathbf{p}_1,\mathbf{p}_2)+
\mathcal{K}^{(1)}_{\alpha\beta}(\mathbf{p}_1,\mathbf{p}_2)+\mathcal{K}^{(2)}_{\alpha\beta}(\mathbf{p}_1,\mathbf{p}_2).
\ee

Distortion and absorption effects are contained in the kernels $\mathcal{K}^{(1)}_{\alpha\beta}$ and $K^{(2)}_{\alpha\beta}$, respectively. The ISI/FSI kernels 
act to remove the probability flux in the excluded overlap region from the free-space probability flux described by the PW kernel. The result is a strongly reduced reaction probability that is confined to the regions of space and momentum that are less affected by the flux absorption. In this formalism, the real parts of the optical potentials, including the long-range ion--ion Coulomb potential, mainly contribute a real-valued phase, i.e., leading to $h_{ij}=e^{i\Phi}|h_{ij}|$.

Changing to $\{\mathbf{P},\mathbf{q}\}$ coordinates, the first-order absorption kernel is found as
\bea
\mathcal{K}^{(1)}_{\alpha\beta}(\mathbf{P},\mathbf{q})=&-&
\int \frac{d^3k_\gamma}{(2\pi)^3}g^{(+)}_{\alpha\gamma}(k_\gamma)
\bigg(\delta(\mathbf{P}+\mathbf{q}/2+\mathbf{q}_{\alpha\gamma})H_S(\mathbf{P}-\mathbf{q}/2+\mathbf{q}_{\gamma\beta})\nonumber\\
&+&\delta(\mathbf{P}-\mathbf{q}/2+\mathbf{q}_{\gamma\beta})\widetilde{H}_S(\mathbf{P}+\mathbf{q}/2+\mathbf{q}_{\alpha\gamma})
\bigg).
\eea

Performing the $k_\gamma$-integration results in
\bea
\mathcal{K}^{(1)}_{\alpha\beta}(\mathbf{P},\mathbf{q})=&-& \frac{1}{(2\pi)^3}
\bigg(H_S(2\mathbf{P}+\mathbf{q}_{\alpha\beta})\nonumber
g^{(+)}_{\alpha\gamma}(\mathbf{P}+\mathbf{q}/2+\mathbf{k}_\alpha|)\\
&+&\widetilde{H}_S(2\mathbf{P}+\mathbf{q}_{\alpha\beta})
g^{(+)}_{\alpha\gamma}(\mathbf{P}-\mathbf{q}/2-\mathbf{k}_\beta|)
\bigg).
\eea

Considering that absorption form factors have a pronounced maximum at $|\mathbf{x}|=0$, we may replace
\be
\mathcal{K}^{(1)}_{\alpha\beta}(\mathbf{P},\mathbf{q})\approx - \frac{1}{(2\pi)^3}
\left(H_S(2\mathbf{P}+\mathbf{q}_{\alpha\beta})
+\widetilde{H}_S((2\mathbf{P}+\mathbf{q}_{\alpha\beta})\right)
g^{(+)}_{\alpha\gamma}(|\mathbf{P}_{\alpha\beta}+\mathbf{q}/2|).
\ee

Using $H_S=e^{2i\Phi}F_S$ and $\widetilde{H}_S=e^{2i\widetilde{\Phi}}F_S$, we obtain
\be
\mathcal{K}^{(1)}_{\alpha\beta}(\mathbf{P},\mathbf{q})\approx - \frac{2}{(2\pi)^3}
e^{i\phi}\cos{(\phi)}F_S(2\mathbf{P}+\mathbf{q}_{\alpha\beta})
g^{(+)}_{\alpha\gamma}(|\mathbf{P}_{\alpha\beta}+\mathbf{q}/2|),
\ee
with $\phi=\frac{1}{2}\left(\Phi+\widetilde{\Phi}\right)$, and we note that the phases may depend on the momentum.

The second-order kernel is
\bea\label{eq:K2}
\mathcal{K}^{(2)}_{\alpha\beta}(\mathbf{P},\mathbf{q})=\int \frac{d^3k_\gamma}{(2\pi)^3}g^{(+)}_{\alpha\gamma}(k_\gamma)
\widetilde{H}_S(\mathbf{P}+\mathbf{q}/2+\mathbf{q}_{\alpha\gamma})
H_S(\mathbf{P}-\mathbf{q}/2+\mathbf{q}_{\gamma\beta}).
\eea

The product of form factors is evaluated easily
by going back to the definition of the form factors $F_S$ as Fourier--Bessel transforms and interchanging the order of momentum and radial integrations. The detailed description in Appendix \ref{app:FSFS} leads to the conclusion that, to a good approximation, the kernel separates into a product of $P$- and $q$-dependent form factors:
\be
\mathcal{K}^{(2)}_{\alpha\beta}(\mathbf{P},\mathbf{q})\simeq
F^{(2)}_{\alpha\beta}(2\mathbf{P}+\mathbf{q}_{\alpha\beta})
g^{(+)}_{\alpha\gamma}(|\mathbf{P}_{\alpha\beta}+\frac{1}{2}\mathbf{q}|).
\ee
where
\be
F^{(2)}_{\alpha\beta}(2\mathbf{P}+\mathbf{q}_{\alpha\beta})=
\int d^3r e^{i(2\mathbf{P}+\mathbf{q}_{\alpha\beta})\cdot \mathbf{r}}
\widetilde{H}_{S}(\mathbf{r})H_{S}(\mathbf{r}).
\ee

Collecting results, we obtain the total reaction kernel:
\bea
&&\mathcal{K}_{\alpha\beta}(\mathbf{P},\mathbf{q})=g^{(+)}_{\alpha\gamma}(|\mathbf{P}_{\alpha\beta}+\frac{1}{2}\mathbf{q}|)\\
&&\times \left(
\frac{1}{8}\delta(\mathbf{P}+\mathbf{q}_{\alpha\beta}/2)-\frac{e^{i\phi}}{(2\pi)^3}\left(
2\cos{(\phi)}F_S(2\mathbf{P}+\mathbf{q}_{\alpha\beta})-e^{i\phi}\overline{F}^{(2)}_{S}(2\mathbf{P}+\mathbf{q}_{\alpha\beta})\right)
\right)\nonumber.
\eea

A phase factor was extracted from the last term: $\overline{F}^{(2)}_{S}\equiv e^{-2i\phi}F^{(2)}_{\alpha\beta}$. The factorized $\mathbf{q}$ and $\mathbf{P}$ dependence is still maintained.

We introduce the ISI/FSI distortion form factor:
\be
\mathcal{D}_{\alpha\beta}(\mathbf{P},\mathbf{q})=\frac{e^{i\phi}}{(2\pi)^3}\left(2\cos{(\phi)}F_S(2\mathbf{P}+\mathbf{q})
-e^{i\phi}\overline{F}^{(2)}_{S}(2\mathbf{P}+\mathbf{q})  \right),
\ee
and performing the $P$ and $q$ integrations, we obtain as the main result of this section the DSCE reaction amplitudes as:
\bea
&&M^{(2)}_{\alpha\beta}(\mathbf{k}_\alpha,\mathbf{k}_\beta))=\\
&&\int d^3q g^{(+)}_{\alpha\gamma}(|\frac{1}{2}\mathbf{q}+\mathbf{P}_{\alpha\beta}|)
\bigg(\frac{1}{8}\mathcal{F}_{\alpha\beta}(-\frac{1}{2}\mathbf{q}_{\alpha\beta},\mathbf{q})
-\int d^3P \mathcal{F}_{\alpha\beta}(\mathbf{P},\mathbf{q}) \mathcal{D}_{\alpha\beta}(\mathbf{P},\mathbf{q}_{\alpha\beta})\bigg. )\nonumber
\eea

\subsection{Momentum Structure of the Nuclear Matrix 
	Element}
The DSCE-NME defined by
\bea
&&\mathcal{F}_{\alpha\beta}(\mathbf{p}_1,\mathbf{p}_2)=\\
&&\sum_{1\neq 3\in \{a\}}\sum_{2\neq 4\in \{A\}}
\lan bB|e^{i\mathbf{p}_2\cdot (\mathbf{r}_3-\mathbf{r}_4)}\mathcal{T}_{NN}(\mathbf{p}_2|34)\mathcal{T}_{NN}(\mathbf{p}_1|12)
e^{i\mathbf{p}_1\cdot (\mathbf{r}_2-\mathbf{r}_2)}|aA\ran \nonumber,
\eea
deserves more detailed consideration because
care has to be taken to avoiding overcounting. This means excluding that the same nucleon takes part in  the first and the second SCE events. Most elegantly, this is achieved by using the coordinates $\{\mathbf{P},\mathbf{q}\}$ and center and relative spatial coordinates $\{\mathbf{r}_\mu,\mathbf{x}_{\mu}\}$, $\mu =(13)$ or $\mu =(24)$, respectively. Hence, we replace $\mathbf{p}_{1,2}=\mathbf{P}\pm \mathbf{q}/2$ and
$r_{1,3}=\mathbf{r}_{13}/2\pm \mathbf{x}_{13}$, and accordingly,
$r_{2,4}=\mathbf{r}_{24}/2\pm \mathbf{x}_{24}$. Furthermore,
for the sake of focusing on the essential features and a more transparent formulation, the momentum dependence of the T-matrices is approximated by $\mathbf{p}_{1,2}\sim \mathbf{q}_{\alpha\beta}/2$, complying with $\mathbf{p}_{1}+\mathbf{p}_{2}\simeq \mathbf{q}_{\alpha\beta}$. Thus, we introduce
\be
U_{NN}(\mathbf{q}_{\alpha\beta}|12,34)=
T_{NN}(\mathbf{p}_2|34)T_{NN}(\mathbf{p}_1|12)_{|_{\mathbf{p}_1=\mathbf{p}_2=\mathbf{q}_{\alpha\beta}/2}}.
\ee

Then, the exclusion principle is treated properly by using
\bea\label{eq:Fab_qab}
&&\mathcal{F}_{\alpha\beta}(\mathbf{P},\mathbf{q})\approx \\
&&4\sum_{1, 3\in \{a\}}\sum_{2, 4\in \{A\}}
\lan bB|e^{i\mathbf{P}\cdot (\mathbf{r}_{13}-\mathbf{r}_{24})}
\mathcal{U}_{NN}(\mathbf{q}_{\alpha\beta}|12,34)
\sin{(\mathbf{q}\cdot \mathbf{x}_{13})}\sin{(\mathbf{q}\cdot \mathbf{x}_{24})}|aA\ran \nonumber ,
\eea
which vanishes for $i=j$ and changes sign under exchanges $1\leftrightarrow 3$ and $2\leftrightarrow 4$, respectively.

At this point, it is advantageous to again incorporate the propagator into the nuclear matrix element such that the $q$-integration can be performed:
\be\label{eq:Bab}
\mathcal{B}_{\alpha\beta}(\mathbf{K},\mathbf{P}_{\alpha\beta})=\int d^3q g^{(+)}_{\alpha\gamma}(|\frac{1}{2}\mathbf{q}+\mathbf{P}_{\alpha\beta}|)
\mathcal{F}_{\alpha\beta}(\mathbf{K},\mathbf{q}).
\ee

Thus, the second-order dynamics originally introduced as a reaction dynamical effect have been converted to
a spectroscopic property. The DSCE reaction amplitude is obtained in the compact and intuitive form
\be\label{eq:M2_Bbar}
M^{(2)}_{\alpha\beta}(\mathbf{k}_\alpha,\mathbf{k}_\beta))=\frac{1}{8}
\mathcal{B}_{\alpha\beta}(-\frac{1}{2}\mathbf{q}_{\alpha\beta},\mathbf{P}_{\alpha\beta})
-\int d^3P \mathcal{B}_{\alpha\beta}(\mathbf{P},\mathbf{P}_{\alpha\beta}) \mathcal{D}_{\alpha\beta}(\mathbf{P},\mathbf{q}_{\alpha\beta}).
\ee

The plane-wave matrix element describes the unit DCE interaction probability without ISI/FSI, thus representing the maximal DCE transition strength. The second component corresponds to a fictitious DCE process occurring in the avoided ion--ion overlap volume. As a result, the heavy-ion DCE amplitude is determined by subtraction of the reaction dynamical forbidden matrix element from the unconstrained plane-wave matrix element of unit interaction probability.

\subsection{Propagator-Induced DSCE Correlations}

From Equation  \eqref{eq:Fab_qab}, it is found that the $q$-dependence is separated, as it is completely contained in the product of the two sine functions. The $q$-integral is easily performed by
substituting $\mathbf{k}=\mathbf{q}/2+\mathbf{P}_{\alpha \beta}$. We also introduce
$\mathbf{x}_{\pm}=\mathbf{x}_{13}\pm \mathbf{x}_{24}$. The addition theorems of trigonometric functions yield:
\bea
C(\mathbf{k},\mathbf{P}_{\alpha\beta})&=& 4\sin{((\mathbf{k}-\mathbf{P}_{\alpha\beta})\cdot \mathbf{x}_{13})}\sin{((\mathbf{k}-\mathbf{P}_{\alpha\beta})\cdot \mathbf{x}_{24})}\\
&=&2\left(C_{-}(\mathbf{k},\mathbf{P}_{\alpha\beta})-C_{+}(\mathbf{k},\mathbf{P}_{\alpha\beta})\right),
\eea
where
\be
C_\pm(\mathbf{k},\mathbf{P}_{\alpha\beta})=
\cos(\mathbf{k}\cdot \mathbf{x}_{\pm})\cos(\mathbf{P}_{\alpha\beta}\cdot \mathbf{x}_{\pm})+
\sin(\mathbf{k}\cdot \mathbf{x}_{\pm})\sin(\mathbf{P}_{\alpha\beta}\cdot \mathbf{x}_{\pm}).
\ee

A closer look shows that the $k$-integral serves to project on the monopole component of the integrand. Due to symmetry, 
only the cosine terms will contribute, leading to
\bea
&&\Gamma_{\alpha\beta}(\mathbf{x}_{13},\mathbf{x}_{24})=\\
&&8\int d^3k g^{(+)}_{\alpha\gamma}(k)
\left(2\cos(\mathbf{k}\cdot \mathbf{x}_{-})\cos(\mathbf{P}_{\alpha\beta}\cdot \mathbf{x}_{-})-
2\cos(\mathbf{k}\cdot \mathbf{x}_{+})\cos(\mathbf{P}_{\alpha\beta}\cdot \mathbf{x}_{+})
\right).\nonumber
\eea

Contour integration leads to the coordinate space propagator
\be\label{eq:gx}
g^{(+)}_{\alpha\gamma}(x)=\int \frac{d^3k}{(2\pi)^3}e^{i\mathbf{k}\cdot \mathbf{x}}g^{(+)}_{\alpha\gamma}(k)=\frac{1}{4\pi x}\left(e^{ik_{+}x}-e^{-ik_{-}x} \right),
\ee
and the correlation function becomes
\be\label{eq:Gamma}
\Gamma_{\alpha\beta}(\mathbf{x}_{13},\mathbf{x}_{24})=64\pi^2
\left(\Gamma^{(-)}_{\alpha\beta}(\mathbf{x}_{13},\mathbf{x}_{24})-
\Gamma^{(+)}_{\alpha\beta}(\mathbf{x}_{13},\mathbf{x}_{24})\right),
\ee
where
\be
\Gamma^{(\pm)}_{\alpha\beta}(\mathbf{x}_{13},\mathbf{x}_{24})=
g^{(\pm)}_{\alpha\beta}(x_{\pm})\cos(\mathbf{P}_{\alpha\beta}\cdot \mathbf{x}_{\pm}).
\ee

The two kinds of correlation functions lead to 
the partial NME:
\bea\label{eq:Bab_Correlated}
&&\mathcal{B}^{(\pm)}_{\alpha\beta}(\mathbf{K}|\mathbf{k}_{\alpha},\mathbf{k}_{\beta})=\\
&&\mp\sum_{1, 3\in \{a\}}\sum_{2, 4\in \{A\}}
\lan bB|e^{i\mathbf{K}\cdot (\mathbf{r}_{13}-\mathbf{r}_{24})}
\mathcal{U}_{NN}(\mathbf{q}_{\alpha\beta}|12,34)
\Gamma^{(\pm)}_{\alpha\beta}(\mathbf{x}_{13},\mathbf{x}_{24})|aA\ran .\nonumber
\eea

The DSCE-NME is obtained as the sum of the two partial matrix elements:
\be\label{eq:BabS}
\mathcal{B}_{\alpha\beta}(\mathbf{K}|\mathbf{k}_{\alpha},\mathbf{k}_{\beta})=\sum_{s=\pm}
\mathcal{B}^{(s)}_{\alpha\beta}(\mathbf{K}|\mathbf{k}_{\alpha},\mathbf{k}_{\beta}).
\ee

Besides the dependencies on the momenta, the properties of $\Gamma^{(\pm)}_{\alpha\beta}$ and
$\mathcal{B}^{(\pm)}_{\alpha\beta}$, respectively, are determined by the propagators $g^{(+)}_{\alpha\beta}(x_{\pm})$.
Their properties are defined by the momenta $k_{0}=\sqrt{\omega^2_\alpha-\omega^2_\gamma}$ and
$\tilde{k}_{0}=-i\sqrt{\omega^2_\alpha+\omega^2_\gamma}$, which are 
determined by the poles of $g^{(+)}_{\alpha\beta}(k)$ in the upper and lower complex half-planes, respectively. Only $k_0$ will be real-valued as long as the auxiliary energy $\omega_\gamma \sim M_\gamma + \bar{\varepsilon}_\gamma$ is less than the c.m. energy $\omega_\alpha$, which is the case when moderate (mean) excitation energies $\bar{\varepsilon}_\gamma$ are 
included in the auxiliary energy $\omega_\gamma$.

\section{Transformation of the DSCE Interactions to S-Channel Form and Nuclear Matrix Elements}\label{sec:DSCE_NME}
\subsection{The DSCE Isotensor Interaction}
The closure approximation has led to effective two-body interactions $\mathcal{R}_{NN}$ acting in the projectile and target nuclei and corresponding, in total, to a nucleus--nucleus four-body interaction. As indicated in Figure \ref{fig:DSCE_prop} by the ladder diagram, originally, the interactions carry an operator structure defined by the t-channel formalism as appropriate for a nuclear reaction. Nuclear matrix elements, however, are defined within a given nucleus, which requires a transformation of the operators into an s-channel formalism. In Ref.~\cite{Lenske:2021jnr}, that problem was solved on the level of matrix elements. Here, we address that question from the operator side by appropriate reordering of operators and application of angular momentum recoupling techniques.

We introduce the rank-2 isotensor operator:
\be \label{eq:IsoTensor}
\mathcal{I}_2(13|24)=\left[\tau_\pm(1)\tau_{mp}(3)\right]_{|a}\left[\tau_\mp(2)\tau_{pm}(4)\right]_{|A}
\ee
and
define the isospin-reduced four-body operators:
\be \label{eq:redRNN}
R_{NN}(\mathbf{p}_1,\mathbf{p}_2|13,24)=S_{NN}(\mathbf{p}_2|34)\otimes S_{NN}(\mathbf{p}_1|12) .
\ee

The effective rank-2 isotensor interaction is then rewritten as
\be\label{eq:RNN}
\mathcal{R}_{NN}(\mathbf{p}_1,\mathbf{p}_2|13,24)=R_{NN}(\mathbf{p}_1,\mathbf{p}_2|13,24)\mathcal{I}_2(13|24).
\ee

Since the isotensor is already in s-channel form, in the following, we need to consider only the reduced DSCE interaction $R_{NN}$.

From the operator structure of the NN T-matrix, we immediately derive that $R_{NN}$ is a superposition of the products of spin scalar, spin vector and spin tensor interactions. The central interactions, for example, lead to the structure:
\be
R^{(0)}_{NN}(\mathbf{p}_1,\mathbf{p}_2|13,24)=\sum_{S,S'=0,1}[\bm{\sigma}_3\cdot \bm{\sigma}_4]^{S'}U_{SS'}(p_1,p_2)[\bm{\sigma}_1\cdot \bm{\sigma}_2]^{S},
\ee
and corresponding expressions are recovered for the tensor interaction and the mixed central tensor terms.
The vertex form factors are
\bea
U_{SS'}(p,p')=V_{S'T}(p')V_{ST}(p)\label{eq:USS}\\
U_{T_nT_n}(p,p')=V_{T_nT}(p')V_{T_nT}(p\label{eq:UT_nT_n})\\
U_{ST_n}(p,p')=V_{S'T}(p')V_{Tn}(p')\label{eq:UST_n}\\
U_{T_nS}(p,p')=V_{T_nT}(p')V_{ST}(p')\label{eq:UT_nS}.
\eea

They are playing the role of momentum-dependent effective coupling constants, and they also depend 
on the energy in the NN rest frame.

While for spin scalar interactions, the t- to s-channel transformation is trivial, the  spin-dependent interactions require considerably more involved treatment by explicit angular momentum recoupling. Since also the spacial degrees of freedom have to be treated properly, we define the spin scalar and spin vector operators:
\be
R_0(\mathbf{k}|i)=e^{i\mathbf{k}\cdot \mathbf{r}_i}\mathbf{1}_{\sigma} \quad ; \quad
\vec{R}_1(\mathbf{k}|i)=e^{i\mathbf{k}\cdot \mathbf{r}_i}\bm{\sigma}_i,
\ee
where $\mathbf{1}_\sigma$ is the spin unity operator; $\mathbf{k}=\mathbf{p}$ for $i=1,3$ in the $a$-system, and $\mathbf{k}=-\mathbf{p}'$ for $i=2,4$ in the $A$-system.

The rank-0 central interactions lead to a superposition of Fermi--Fermi (FF), Gamow--Teller--Gamow--Teller (GG) and mixed Fermi--Gamow--Teller transitions (FG and GF). Leaving aside from hereon the vertex form factors, the  resulting operators are
\bea
\Sigma_{FF}(\mathbf{p}_1,\mathbf{p}_2|13,24)&=&R_0(\mathbf{p}_1|1)R_0(\mathbf{p}_2|3)R_0(\mathbf{p}_1|2)R_0(\mathbf{p}_2|4)\\
\Sigma_{FG}(\mathbf{p}_1,\mathbf{p}_2|13,24)&=&
R_0(\mathbf{p}_2|3)\left[\vec{R}_1(\mathbf{p}_1|1)\cdot\vec{R}_1(\mathbf{p}_1|2)\right]R_0(\mathbf{p}_2|4)\\
\Sigma_{GF}(\mathbf{p}_1\mathbf{p}_2|13,24)&=&
R_0(\mathbf{p}_1|1)\left[\vec{R}_1(\mathbf{p}_2|3)\cdot\vec{R}_1(\mathbf{p}_2|4)\right]R_0(\mathbf{p}_1|2).
\eea

While the FF and FG/GF components are already in an appropriate form, the double Gamow--Teller operators (GG) require a more detailed treatment. The products of spin vector components are recoupled by arranging the nucleon spin operators into intra-nuclear two-body operators of tensorial rank 0, 1 and 2, which by contraction form a total rank-0 scalar operator in the overall projectile--target system:
\be
\Sigma_{GG}(\mathbf{p}_1,\mathbf{p}_2|13,24)=
\sum_{S=0,1,2}
\left[\vec{R}_1,(\mathbf{p}_1|1)\otimes \vec{R}_1(\mathbf{p}_2|3)\right]_{S}\cdot
\left[\vec{R}_1,(\mathbf{p}_1|2)\otimes \vec{R}_1(\mathbf{p}_2|4)\right]_S
\ee
where the $S=0$ components are, in fact, scalar products: $\vec{R}_1(\mathbf{p}_1|i)\cdot\vec{R}_1(\mathbf{p}_2|j)$.

An even richer spectrum of operators is obtained from the double-tensor (TT) term. Labeling for bookkeeping reasons the nucleon spin operators by $S_i$, where $S_i=1$, defining $S_{ij}=2$, using $\widehat{J}=\sqrt{2J+1}$, and applying angular momentum recoupling techniques resulting in 9-j symbols, one finds
\bea
&&\Sigma_{T_nT_n}(\mathbf{p}_2,\mathbf{p}_2|13,24)=
\widehat{S}_{12}\widehat{S}_{34}\sum_{S_a,S_A=0,1,2}\widehat{S}_a\widehat{S}_A \sum_{S=|S_a-S_A|}^{S_a+S_A}
\left\{
  \begin{array}{ccc}
    S_1 & S_2 & S_{12} \\
    S_3 & S_4 & S_{34} \\
    S_a & S_A & S \\
  \end{array}
\right\}
\nonumber\\
&&\times \left[Y_{S_{34}}(\widehat{\mathbf{p}}')\otimes Y_{S_{12}}(\widehat{\mathbf{p}}) \right]_{S}\cdot \left[
\left[\vec{R}_1(\mathbf{p}|1)\otimes \vec{R}_1(\mathbf{p}'|3)\right]_{S_a}\otimes
\left[\vec{R}_1(\mathbf{p}|2)\otimes \vec{R}_1(\mathbf{p}'|4)\right]_{S_A}\right]_{S}.
\eea

The mixed central-tensor terms include the Fermi--Tensor components FT and TF:
\bea
&&\Sigma_{FT}(\mathbf{p}_1,\mathbf{p}_2|13,24)=
Y_2(\widehat{\mathbf{p}}_1)\cdot \left[R_0(\mathbf{p}_2|3)\vec{R}_1(\mathbf{p}_1|1)\otimes \vec{R}_1(\mathbf{p}_1|2)R_0(\mathbf{p}_2|4)\right]_2\\
&&\Sigma_{TF}(\mathbf{p}_1,\mathbf{p}_2|13,24)=
Y_2(\widehat{\mathbf{p}}_2)\cdot \left[R_0(\mathbf{p}_1|1)\vec{R}_1(\mathbf{p}_2|3)\otimes \vec{R}_1(\mathbf{p}_2|4)R_0(\mathbf{p}_1|2)\right]_2 .
\eea

With appropriate recoupling, the combined Gamow--Teller--Tensor (GT) components become
\bea
&&\Sigma_{GT}(\mathbf{p}_1\mathbf{p}_2|13,24)=\widehat{L}^2\sum_{S_a,S_A=1,2}\widehat{S}_a\widehat{S}_A
\left\{
  \begin{array}{ccc}
    S_1 & S_2 & L \\
    S_3 & S_4 & 0      \\
    S_a & S_A & L \\
  \end{array}
\right\} \nonumber \\
&&\times Y_{L}(\widehat{\mathbf{p}})\cdot
\left[\left[\vec{R}_1(\mathbf{p}_1|1)\otimes \vec{R}_1(\mathbf{p}_2|3)\right]_{S_a}\otimes
      \left[\vec{R}_1(\mathbf{p}_1|2)\otimes \vec{R}_1(\mathbf{p}_2|4)\right]_{S_A}\right]_{L} ,
\eea
with $L=2$. Accordingly,
\bea
&&\Sigma_{TG}(\mathbf{p}_1,\mathbf{p}_2|13,24)=\widehat{L}^2\sum_{S_a,S_A=1,2}\widehat{S}_a\widehat{S}_A
\left\{
  \begin{array}{ccc}
    S_1 & S_2 & 0      \\
    S_3 & S_4 & L \\
    S_a & S_A & L \\
  \end{array}
\right\}
\nonumber \\
&&\times Y_{L}(\widehat{\mathbf{p}}_2)\cdot
\left[\left[\vec{R}_1(\mathbf{p}_1|1)\otimes \vec{R}_1(\mathbf{p}_2|3)\right]_{S_a}\otimes
      \left[\vec{R}_1(\mathbf{p}_1|2)\otimes \vec{R}_1(\mathbf{p}_2|4)\right]_{S_A}\right]_{L}.
\eea
By expanding the plane wave parts of the spin--scalar and spin--vector operators a rich spectrum of multipole operators is obtained with radial form factors given by Riccati--Bessel functions and spherical harmonics accounting for the orbital angular momentum transfer. Details of the expansion and the definition of the resulting coupled  multipole operators are discussed in appendix \ref{app:SpinScalar} and in appendix \ref{app:SpinVector} for spin--scalar and spin--vector operators, respectively. In appendix \ref{app:BiSpherical} the frequently occurring bi--spherical harmonics are investigated.

\subsection{DSCE Form Factors and Spectroscopy}
The DSCE transition form factor given by
\be
\mathcal{F}_{\alpha\beta}(\mathbf{p}_1,\mathbf{p}_2)=\sum_{ij\in \{a\},kl\in \{A\}}\lan bB|\mathcal{R}_{NN}(\mathbf{p}_1,\mathbf{p}_2|ij,kl)|aA\ran
\ee
generalizes the concept of nuclear matrix elements to transition form factors at finite momentum transfers. Nuclear matrix elements in the strict sense are recovered for  $p,p' \to 0$, which is 
known as the \emph{long wavelength 
limit}~\cite{FeshbachDeShalit:1974nuc} and is widely used in the electro-weak sector.

A key question is what kind of spectroscopic information can be extracted from DSCE cross sections. For an answer, we have to have a closer look into the operator structure, which is given by a
multitude of terms. Already from the central spin vector interactions, two additional spin tensor terms were obtained by recoupling. The majority of components are generated by the additional degrees of freedom provided by the rank-2 spin tensor interactions. Counting the  terms obtained by recoupling separately, more than 40 components are identified. However, they are determined by four generic types of operators: namely, the spin scalar--scalar terms, mixed scalar--vector and scalar--tensor, and rank-1 and rank-2 spin tensor terms. The scalar $R_0$ and vector $\mathbf{R}_1$ operators contain a rich spectrum of multipole operators, which are obtained by expanding the plane-wave factors into partial waves, as discussed in Appendices \ref{app:SpinScalar} and \ref{app:SpinVector}, respectively.

As an example, we consider $\Sigma_{FG}$ and introduce the partial form factor
\be
F^{(FG)}_{\alpha\beta}(\mathbf{p},\mathbf{p}')=\sum_\mu(-)^\mu F^{FG}_{ab,\mu}(\mathbf{p},\mathbf{p}')F^{FG}_{AB,-\mu}(\mathbf{p},\mathbf{p}')
\ee
where the isospin matrix elements were left out for simplicity, and the scalar product is evaluated explicitly in the spherical bi-orthogonal basis $\{\mathbf{e}^*_\mu,\mathbf{e}_\mu\}$, $\mu=0,\pm 1$; see Appendix~\ref{app:SpinVector}. The form factors in the $a$- and the $A$-systems, respectively, are defined as:
\bea
\vec{F}^{(FG)}_{ab}(\mathbf{p},\mathbf{p}')&=&\sum_{(13)}\lan b|\vec{R}_1(\mathbf{p}'|3)R_0(\mathbf{p}|1)|a\ran
=\sum_{\mu} F^{(FG)}_{ab,\mu}(\mathbf{p},\mathbf{p}')\mathbf{e}^*_\mu\\
\vec{F}^{(FG)}_{AB}(\mathbf{p},\mathbf{p}')&=&\sum_{(24)}\lan B|\vec{R}_1(\mathbf{p}'|4) R_0(\mathbf{p}|2)|A\ran
=\sum_{\mu} F^{(FG)}_{AB,\mu}(\mathbf{p},\mathbf{p}')\mathbf{e}^*_\mu.
\eea

Both form factors are given by the products of spin scalar and spin vector operators---the latter being expressed in spherical representation following the rules discussed in the appendix.

The DSCE nuclear form factors and NMEs are given by a superposition of terms, which are factorized into a projectile and a target NME. However, in general, complete separation into a single product of nuclear NMEs is rather unlikely, even for $0^+ \to 0^+$ DCE transitions in both reaction partners. As an example, we consider the $A(0^+)\to B(0^+)$ case. Obviously, the total angular momentum transfer is restricted to $J^\pi=0^+$. However, by a two-body operator, that transition can be achieved in at least two ways: namely, by the total monopole part of the FF operator $\left[R_0(\mathbf{p}|1)\otimes R_0(\mathbf{p}'|3)\right]_{J=0}$ or by the coupling of the total orbital/spin quadrupole components of the GG-operator coupled to a monopole operator:
$\left[\mathbf{R}_1(\mathbf{p}|1)\otimes \mathbf{R}_1(\mathbf{p}'|3)\right]_{(L=S=2)J=0}$. Excitations of final states of higher angular momentum will enlarge the number of allowed contributions considerably.

\subsection{Direct Evaluation of the Nuclear Matrix Elements}
Although the investigations in the foregoing section are of high value to understand the dynamics of DCE transitions induced by NN interactions, they may not be the most favorable approach for practical numerical calculations. Moreover, the formalism seems to be quite different from the one derived in Ref.~\cite{Lenske:2021jnr}, which led to rank-2 polarization tensors resembling $2\nu 2\beta$-NME. In the following, we show that, in fact, the present and former results are in perfect agreement.

Going back to Equation  \eqref{eq:redRNN}, we may separate the product of scattering operators by inserting the complete set of intermediate states at the proper place and find:
\bea
\mathcal{F}_{\alpha\beta}(\mathbf{p},\mathbf{p}'|13,24)=\sum_{c,C}\lan bB|\mathcal{S}_{NN}(\mathbf{p}'|34)|cC\ran\cdot \lan cC|\mathcal{S}_{NN}(\mathbf{p}|12)|aA\ran,
\eea
thereby reversing closure by re-installing the spectrum of intermediate states. Contraction to a total scalar is indicated by the dot-product. With the presentation of $\mathcal{S}_{NN}$ by products of the one-body spin scalar and spin vector operators $R_0(\mathbf{p}|i)$ and $\mathbf{R}_1(\mathbf{p}|i)$, respectively, we obtain
\be
\mathcal{F}_{\alpha\beta}(\mathbf{p},\mathbf{p}'|13,24)=\sum_{S_1,S_2}U_{S_1S_2}(p,p')\cdot\mathcal{F}_{S_1S_2}(\mathbf{p},\mathbf{p}'|ab)
\mathcal{F}_{S_1S_2}(\mathbf{p},\mathbf{p}'|AB),
\ee
and the spin tensor terms are treated analogously.
The projectile and target NMEs are
\bea
\mathcal{F}_{S_1S_2}(\mathbf{p},\mathbf{p}'|ab)&=&
\sum_{c\in \{a\}}\lan b|\mathcal{R}_{S_2}(\mathbf{p}'|3)|c\ran\lan c|\mathcal{R}_{S_1}(\mathbf{p}|1)|a\ran\\
\mathcal{F}_{S_1S_2}(\mathbf{p},\mathbf{p}'|AB)&=&
\sum_{C\in \{A\}}\lan B|\mathcal{R}_{S_2}(\mathbf{p}'|4)|C\ran\lan C|\mathcal{R}_{S_1}(\mathbf{p}|2)|A\ran.
\eea

The relation to the polarization tensor formalism developed in~\cite{Lenske:2021jnr} is seen by rewriting the NME in the form of a contour integral over a rank-2 polarization tensor:
\bea
\mathcal{F}_{S_1S_2}(\mathbf{p},\mathbf{p}'|ab)=
\frac{1}{2i\pi}\oint d\omega \sum_{c\in \{a\}}\lan b|\mathcal{R}_{S_2}(\mathbf{p}'|3)|c\ran\frac{1}{E_c-\omega+i\eta}\lan c|\mathcal{R}_{S_1}(\mathbf{p}|1)|a\ran.
\eea

The NME of the A-system is treated  accordingly.
Performing a multipole expansion as needed for nuclear spectroscopy, the coupling schemes developed in the previous section must be applied. Exploiting the addition theorems for trigonometric and Riccati--Bessel functions, the correlation functions $\Gamma^{(\pm)}_{\alpha\beta}$ can be expanded in products of correlation functions acting in the $a$-like and $A$-like sub-systems. That allows us to incorporate the correlations into the above NME.

\section{Illustrative Applications for DSCE Reactions}\label{sec:IllusDSCE}
\subsection{The Reaction}
The approach discussed in the previous sections has been applied to the DCE reaction $^{40}$Ca$(^{18}$O$,^{18}$Ne$)^{40}$Ar at $T_{lab}=270$~MeV measured at LNS Catania~\cite{Cappuzzello:2015ixp}.
Following the approach discussed in~\cite{Lenske:2018jav}, the absorption radius $R_{abs}\simeq 8.40$~fm was derived from the total reaction cross
section $\sigma_{reac}=2.218$~b in the incident channel. The latter was obtained by an optical model calculation that included partial wave equations up to angular momentum $\ell =200$. A double folding potential was used and was calculated with Hartree--Fock--Bogolyubov (HFB) ground state densities of the A=18 and A=40 nuclei and a NN T-matrix~\cite{,Cappuzzello:2022ton} that was newly derived for NN energies in the region $T_{lab})10\ldots 50$~MeV in Love--Franey parametrization~\cite{Love:1981gb,Franey:1985ye,Love:1987zx}.

\begin{figure}[H]

\includegraphics[width = 13.5cm]{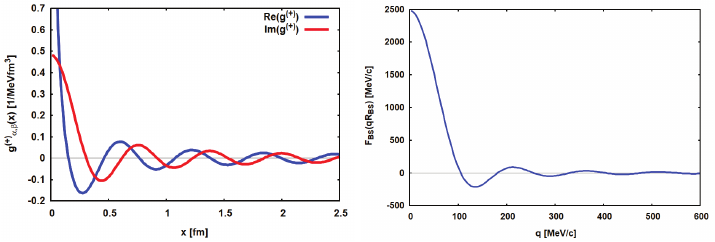}
\caption{The propagator 
 $g^{(+)}_{\alpha\gamma}(x)$ for the reaction $^{40}$Ca$(^{18}$O$,^{18}$Ne$)^{40}$Ar at $T_{lab}=270$~MeV with $k_0\sim 2017$~MeV/c is shown on the left. The real part (blue) and the imaginary part (red) are shown separately.  On the right, the black sphere distribution for the BS radius $R_{abs}=8.40$~fm is displayed. }
\label{fig:gx}

\end{figure}

For that reaction, the propagator contains a real-valued pole at $k_0\simeq 2017$~MeV/c corresponding to a kinetic energy $T_0\simeq 176$~MeV if in the intermediate channel excitation energies are neglected. Adding excitation energy will move the pole to lower values of $k_0$, but qualitatively, the same results are obtained. As seen in Figure \ref{fig:gx}, the propagator $g^{(+)}_{\alpha\beta}(x)$ is narrowly peaked at small distances $x$, confirming the mentioned strong localization around $x\sim 0$.

\subsection{The Black Sphere Limit}
Explicit calculations for strongly absorbing systems like interacting heavy ions show that the shape of the reduced distortion form factors $h_{ij}=1-\eta_{ij}$ resembles a Heaviside distribution $|h_{ij}(\mathbf{r})|\simeq  h_B(r)=\Theta(R_{abs}-r)$, where $h_B(x)$ describes the excluded volume of the absorptive overlap region, known as the \emph{black sphere} (BS) approximation. In momentum space, the BS reaction kernels are obtained by the Fourier--Bessel transform of $h_B(r)$,  leading to the form factor $h_B(x)=3j_1(x)/x$, $h_B(0)=1$, where $x=kR_{abs}$ and $j_1(x)$ is a spherical Riccati--Bessel function. In BS approximation, the reduced absorption form factor is given by
\be
H_S(\mathbf{k})\mapsto F_{BS}(\mathbf{k})=\frac{R^3_{abs}}{6\pi^2}h_B(kR_{abs}).
\ee

The function $F_{BS}$ is strongly peaked at $x=0$ with a width $\simeq 1/R_{abs}$. Since $R_{abs}(A_P,A_T)$ increases with the mass numbers, the width of $F_{BS}$ decreases with increasing $A_{P,T}$, approaching as the limiting case $F_{BS}$ a delta distribution.

We also note that in the black sphere limit, the second-order form factor becomes
\be
F^{(2)}_{\alpha\beta}(\mathbf{P}|q_{\alpha\beta})\approx \frac{(R^{(2)}_{abs})^3}{6\pi^2} e^{i\phi}h_B(|\mathbf{P}+\mathbf{q}_{\alpha\beta}|R^{(2)}_{abs}).
\ee

For $R^{(2)}_{abs}=R_{abs}$, the kernel attains an intriguing simple form:
\be
\mathcal{K}_{\alpha\beta}(\mathbf{P},\mathbf{q})\to  g^{(+)}_{\alpha\gamma}(|\mathbf{P}_{\alpha\beta}+\frac{1}{2}\mathbf{q}|)
\left(
\delta(\mathbf{P}+\mathbf{q}_{\alpha\beta})-\frac{1}{6\pi^2}\left(\frac{R_{abs}}{2\pi}\right)^3h_B(|\mathbf{P}+\mathbf{q}_{\alpha\beta}|R_{abs})
\right)
\ee

The minus sign indicates the sizable reduction in the DSCE cross section by several orders of magnitude due to cancellation of the plane-wave part by the absorption exerted by the imaginary part of the ion--ion optical potential. In the following case studies, the BS approximation will be used, mainly because of its especially transparent structure and easy reproducibility.

\subsection{Induced Correlations}
An important ingredient of DSCE dynamics is the previously introduced correlation functions $\Gamma^{(\pm)}_{\alpha\beta}$. Since they are controlling the transition form factors, their properties are decisive for the observed NME and, moreover, for the whole reaction.
As seen in Figure \ref{fig:gx}, the propagator encountered in
the reaction $(^{18}$O+$^{40}$Ca at $T_{lab}=270$~MeV has a maximum at $|\mathbf{x}|=0$. At larger distances, the oscillatory pattern declines as $1/x$. Averaging the correlation functions over the orientations of $\mathbf{x}$ amounts to evaluating
\be
\lan \cos{(\mathbf{P}_{\alpha\beta}\cdot \mathbf{x})}\ran = j_0(P_{\alpha\beta}x)
\ee
where
$j_0(x)$ is the zeroth-order Riccati--Bessel function, which oscillates with a period determined by the value of half the sum of the incident and exit channel momenta $\mathbf{P}_{\alpha\beta}$.

For the above reaction, one finds in the measured angular range $|\mathbf{P}_{\alpha\beta}|\sim 2060$~MeV/c, which is 
almost identical to the momentum $k_0=2017$~MeV/c by which the propagator evolves. Hence, both functions oscillate with  periods $\sim 10$~fm$^{-1}$.
In total, one finds that $\Gamma^{(\pm)}_{\alpha\beta}$ are strongly peaked at small values of $x_\pm$. The short-range character is confirmed by the root mean square (rms) radius of the correlation function
$\lan x^2 \ran^\frac{1}{2}\simeq 1.96$~fm evaluated over a volume with a radius of half the distance between the centers of $^{18}$O and $^{40}$Ca. The rms value is in surprisingly good agreement with the two-nucleon correlation $x^{(0\nu)}_{NN}=1\ldots 4$~fm induced by the exchange of the Majorana neutrinos, as found in $0\nu 2\beta$ DBD studies. Thus, $\Gamma_{\alpha\beta}$ favors spatial configurations with $\mathbf{x}_{13}\sim \pm \mathbf{x}_{24}$, implying that the pairs of SCE vertices in the projectile and target are arranged in almost the same manner.
The correlations, however, are of a rather fragile character. In addition to the incident energy and the nuclear masses, 
they also depend on the scattering angle through $\mathbf{P}_{\alpha\beta}$ and on the (mean) intermediate excitation energy contained in $\omega_\gamma$.

\subsection{The DSCE Cross Section in DW and BS Approximation}

\begin{figure}[H]

\includegraphics[width = 8.8cm]{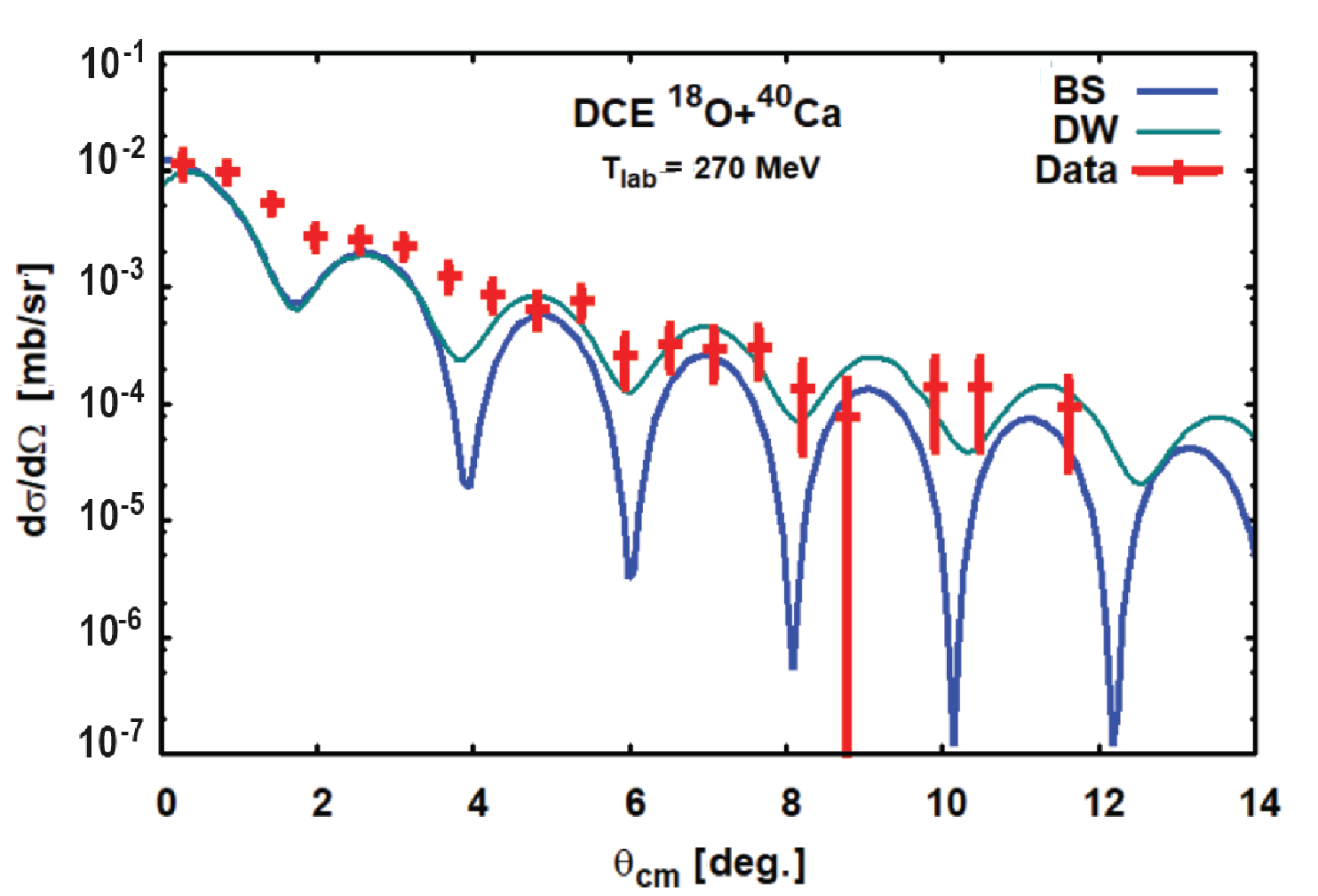}
\caption{Second
-order black sphere ( 
BS, bold blue line) results for the reaction $^{40}$Ca$(^{18}$O$,^{18}$Ne$)^{40}$Ar at $T_{lab}=270$~MeV are compared to data~form \cite{Cappuzzello:2015ixp} and the full second-order DW  results from Ref.~\cite{Bellone:2020lal} (DW, grey thin line). The  BS results are normalized to the DW results by a $\chi^2$ fit.}
\label{fig:DSCE_BS}

\end{figure}

In Figure \ref{fig:DSCE_BS}, the results obtained with the second-order black sphere model are displayed and compared to the full second-order DW results of~\cite{Bellone:2020lal} and the measured angular distribution. In~\cite{Bellone:2020lal}, second-order DW calculations were performed in a fully microscopic approach. The optical potentials were calculated in a double folding approach using HFB proton and neutron ground state densities and the isoscalar and isovector parts of the NN T-matrix. Also the ion--ion Coulomb potentials were obtained microscopically by double folding the nuclear charge densities with the two-body Coulomb interaction, including contributions due to anti-symmetrization; see, e.g.,~\cite{Satchler:1983,Feshbach:2003,Lenske:2021bpk}. The ion--ion potentials were checked against elastic scattering data. The optical model wave equations were integrated numerically for partial waves up to $\ell \sim 200$, ensuring convergence for elastic and SCE and DCE angular distributions. The second-order DSCE reaction amplitude was constructed according to Equation  \eqref{eq:MDSCE_M1M1} but in pole approximation. The first-order SCE-type reaction amplitudes were calculated in partial wave representation and covered 
nuclear transitions with multipolarities $J^\pi=0^\pm\ldots 5^\pm$ in the projectile and target. The nuclear transition form factors and response functions were obtained in QRPA calculations using the Giessen energy density functional~\cite{Tsoneva:2017kaj,Lenske:2019ubp,Adamian:2021gnm}. QRPA spectral distributions are found in~\cite{Lenske:2018jav}. It is worth mentioning that the measured angular distribution is reproduced in absolute terms without additional adjustments. Note that the angular range covered by the data corresponds to a remarkable range of momentum transfers up to $q_{\alpha \beta}\sim 400$~MeV/c.

Following Equation \eqref{eq:M2_Bbar}, the BS reaction amplitude is constructed by superimposing the plane-wave amplitude, which contains the DSCE-NME without ISI/FSI, and the distortion amplitude, which accounts 
for ISI/FSI. In~\cite{Bellone:2020lal}, second-order PW and second-order DW amplitudes were compared; the study impressively showed the anticipated strong suppression of the strength by about five orders of magnitude. In BS approximation, the same effect is produced by the interfering distortion amplitude. The DCE reaction is a $0^+\to 0^+$ transition in both nuclei that constrains 
the total angular momentum transfer to $J=0$. However, as discussed in Section \ref{sec:DSCE_NME}, this is compatible to two combinations of total orbital and spin transfers, $L=0,S=0$ and $L=2,S=2$, where $\mathbf{L}=\mathbf{L}_1+\mathbf{L}_2$ and $\mathbf{S}=\mathbf{S}_1+\mathbf{S}_2$ are the properly coupled total angular momentum and spin transfers in the first and the second step of the DCE reaction. Hence, the NME  is given by an $L=0,S=0$ and an $L=2,S=2$ amplitude, where the monopole component dominates by far at small scattering angles. The PW amplitude, i.e., the ISI/FSI-free NME, is well-described by a zeroth and second-order Riccati--Bessel function with the same radius parameter $R_{PW}\simeq 7.8$~fm. For the ISI/FSI amplitude, we find $R_{abs}\simeq 9.0$~fm. The two kinds of amplitudes interfere with the phase $\phi\simeq 172.8^\circ$ that was fitted to the data. Hence, the PW and ISI/FSI contributions interfere almost destructively, which explains the strong suppression of the transition strength.

As anticipated in Section \ref{ssec:DisAbs}, the ISI and FSI terms  not only induce strong suppression of the DSCE amplitude with respect to the plane-wave limit but also imprint on the DCE angular distributions their own diffraction structure caused by ISI and FSI and reflecting the size of the excluded overlap volume of strong absorption. The suppression effect and the modification of the diffraction structure by ISI and FSI were already illustrated in~\cite{Bellone:2020lal}, where second-order PWA and DWA angular distributions were compared.
Thus, the diffraction pattern observed for heavy-ion DCE cross sections is mainly determined by the optical model potentials and only to a lesser degree by the transition form factors. The multipolarity of the form factors, however, determines the overall shape of an angular distribution, especially their behavior during small momentum transfers.
That effect was already discussed in~\cite{Lenske:2018jav} for SCE reactions, but in the DCE case, the ISI/FSI suppression is considerably enhanced.

In view of the simplicity of the black sphere model, the overall agreement between the two model calculations and the data is remarkable. At forward angles, i.e., small momentum transfers, the two theoretical approaches give almost identical results. The deviations, which develop with increasing scattering angles, indicate---as to be expected---the remaining differences of the schematic, semi-classical black sphere model to the quantal second-order DW calculation.

The question may arise as to what extent the BS results depend on the choice of the functional form of the distortion form factors.
That point was checked by alternatively using form factors with a Fermi-function shape with a diffuse surface: 
thus modeling an opaque \emph{gray sphere}. However, to reproduce with satisfactory accuracy the distortion coefficients obtained with the distorted waves from optical model calculations together with the DCE angular distribution, a rather small diffusivity parameter $\sim 0.1$~fm is required. With such a small value, the Fermi-distribution converges with very good approximation to a Heaviside distribution.

\section{Summary and Discussion}\label{sec:Sum}
Starting from a fully microscopic and quantal approach to a heavy-ion DSCE reaction, the role of initial state (ISI) and final state (FSI) interactions and the induced rank-2 isotensor interaction were investigated. The DSCE amplitude, being an, in principle, well known distorted wave two-step structure, was reconsidered in a momentum space approach that allowed the separation of the ISI and FSI contributions from nuclear matrix elements.  Since heavy-ion DCE reactions are peripheral direct reactions, the elastic ion--ion interactions underlying ISI and FSI are well described by optical model potentials: either of empirical or theoretical origin. The accuracy achieved in the description of DCE data---as in all other kinds of direct nuclear reactions---depends, of course, on the quality of the optical potentials. In order to control that part of the theory, elastic scattering data for at least the incident $a+A$ system are indispensable as a probe for the optical potentials and their optimization.

The critical parts for understanding the induced suppression of transition strengths are the imaginary potentials. They are acting as sinks for the probability flux of the scattering waves by redirecting a large fraction of elastic flux into other reaction channels in a \emph{never come back} manner. Distortions and absorption effects were arranged into reaction kernels. An important aspect of ISI and FSI is that they affect the scattering wave functions. Hence, the relation to the optical potentials is of a complex nature that is mathematically defined by the underlying differential equation. The highly non-linear relation between the potential and wave amplitude is the reason that the absorption amplitudes are determined by the ion--ion total reaction cross sections~\cite{Lenske:2005ntx,Lenske:2018jav}. Total reaction cross sections are the observable consequences of flux absorption by coupled channel dynamics.

In a nuclear DCE reaction, the transition form factors are the relevant nuclear structure objects. Nuclear matrix elements are obtained in the limit of vanishing momentum transfer, known as the long wavelength limit. By approximating the intermediate Green's function by an average, channel-independent propagator, the DCE form factors can be investigated in closure approximation. Different to two-neutrino DBD, the closure approach is well justified for heavy-ion DCE reactions because nuclear interactions excite a large spectrum of intermediate states and multipolarities. detailed study of the interplay between the nuclear structure and reaction effects led to the interesting result that the intermediate propagation induces correlations between the first and second SCE events by imposing constraints on the vertices. ISI and FSI were discussed in the black sphere limit as a simple to handle but realistic approximation. The comparison of BS cross section results to full second-order DSCE calculations and DCE data led to surprisingly good agreement for the angular distribution.

A section was devoted to derive from the pair of two-body NN T-matrices the effective DSCE interaction. That goal was achieved  by transforming the reaction-theoretical t-channel interactions into the s-channel representation required for nuclear structure studies. At the end, the products of t-channel operators were rearranged and recoupled into products of an s-channel operator. The s-channel interaction is given by tensorial dyadic products of two-body operators describing the $n^2p^{-2}$ and the associated complementary $p^2n^{-2}$ DCE transitions of two particle-two hole character in the projectile and target.

The BS model, albeit having convincing simplicity and surprising success, is not meant to replace a full microscopic description of DCE reactions. First of all, an optical model calculation is still needed for the determination of the absorption parameters. At sufficiently high energies---typically reached at about $T_{lab} \sim 100$~AMeV---the distortion and absorption effects can be treated safely by eikonal theory, which will simplify that step considerably. When extracting NMEs, it remains to understand their spectroscopic content. In general, that task will be confronted  by disentangling a superposition of nuclear spin scalar and spin vector multipoles in the projectile and target. Thus, nuclear structure calculations remain an indispensable tool for identifying the spectroscopic content. Moreover,
what we learn from the BS case studies is that the proper treatment of ISI/FSI effects is essential for a realistic description of the magnitude and the diffraction pattern of heavy-ion DCE angular distributions.

As a closing remark, we emphasize that modern nuclear reaction and nuclear structure theory, as used, e.g., in~\cite{Bellone:2020lal}, provides the proper theoretical tools and numerical methods for the quantitative description of processes as complex as a DCE reaction. The scope of this article was to clarify the role of ISI/FSI dynamics and their interplay with residual ion--ion interaction and to elucidate their cooperation for inducing correlations and effective isotensor interactions. It should be noted that the effective DSCE isotensor interaction is a four-body operator. Heavy-ion DCE reactions provide, for the first time, the environment appropriate to investigate  such high-rank operators on a data-driven basis.

\authorcontributions{H.L.: Conceptualization, methodology, original draft preparation and funding acquisition, J.B.: investigation and methodology, M.C.: investigation, administration, supervision, and funding acquisition, D.G.: investigation and formal analysis, J.-A.L.: investigation and funding acquisition. All authors have read and agreed to the published version of the manuscript.} 

\funding{H. Lenske acknowledges
financial support in part by DFG, grant Le439/16-2, and INFN/LNS Catania. J.-A. Lay acknowledges that this work is based on research supported
in part by grant No. PID2020-114687GB-I00 funded by MCIN/AEI/10.13039/501100011033.}

\dataavailability{Not applicable} 

\conflictsofinterest{The authors declare no conflicts of interest.}

\appendixtitles{yes} 
\appendixstart
\appendix
\section[\appendixname~\thesection]{Auxiliary Energy and Residual Term  for the Closure Approximation \label{app:Residual}}
In $Res(\xi)$, the residual terms beyond closure are collected. The next-to-leading-order term, i.e., the leading-order term in $\xi$, is
\bea
Res(\xi)=
\left(
 \frac{\omega_\gamma}{\left(\omega_\alpha -\omega_\gamma -T_\gamma(k_\gamma) +i\eta\right)^{2}}
-\frac{\omega_\gamma}{\left(\omega_\alpha +\omega_\gamma +T_\gamma(k_\gamma) +i\eta\right)^{2}}
\right)\lan\xi\ran + \mathcal{O}(\lan\xi^2\ran)
\eea
which is already suppressed energetically by the quadratic energy denominators. The expected 
value of $\xi$ is
\be\label{eq:LO}
\lan\xi\ran=\mathcal{E}_0(\mathbf{p},\mathbf{p}'|aA)-\frac{1}{\omega_\gamma}\mathcal{E}_1(\mathbf{p},\mathbf{p}'|aA),
\ee
where
\be
\mathcal{E}_n(\mathbf{p},\mathbf{p}'|aA)=
\sum_{cC}
\left(E_c+E_C  \right)^n
\lan bB|\mathcal{S}_{NN}(\mathbf{p}'|34)|cC\ran\lan cC|\mathcal{S}_{NN}(\mathbf{p}|12)|aA\ran
\ee

By closure, we recover for $n=0$ the DSCE-NME:
\be
\mathcal{E}_0(\mathbf{p},\mathbf{p}'|aA)=
\lan bB|\mathcal{S}_{NN}(\mathbf{p}'|34)\mathcal{S}_{NN}(\mathbf{p}|12)|aA\ran
\ee
and for $n=1$, we have the 
sum of a new kind of rank-2 energy-weighted sum rules defined for the transition from the parent nuclei to states in the daughter nuclei:
\bea
\mathcal{E}_1(\mathbf{p},\mathbf{p}'|aA)&=&
\sum_{c\in\{a\}}
E_c\lan bB|\mathcal{S}_{NN}(\mathbf{p}'|34)|c\ran\lan c|\mathcal{S}_{NN}(\mathbf{p}|12)|aA\ran \nonumber\\
&+&\sum_{C\in\{A\}}
E_C\lan bB|\mathcal{S}_{NN}(\mathbf{p}'|34)|C\ran\lan C|\mathcal{S}_{NN}(\mathbf{p}|12)|aA\ran .
\eea

A closer look reveals that this corresponds to the sum of energy-weighted sum rules of products of the spin scalar, spin vector and spin tensor in the projectile and target.

As seen from Equation  \eqref{eq:SNN}, the sum rules include momentum dependence from the plane-wave factors and the (products of) $V_{ST}$ and $V_{Tn}$. However, that dependence is largely canceled when
defining the auxiliary energy by
\be
\omega_{\gamma}\simeq  \frac{\mathcal{E}_1(\mathbf{p},\mathbf{p}'|aA)}{\mathcal{E}_0(\mathbf{p},\mathbf{p}'|aA)}.
\ee

By that choice, the leading-order term in $\xi$ from Equation \eqref{eq:LO} 
is canceled, and the higher-order contributions will be minimized. The auxiliary energy is, in fact, a functional of the reacting nuclei and their intrinsic and mutual interactions: $\omega_\gamma=\omega_\gamma[H_a,H_A,V_{aA}]$.

\section[\appendixname~\thesection]{The Second-Order Reaction Kernel}\label{app:FSFS}
With $\mathbf{r}_\alpha=\mathbf{r}+\mathbf{x}/2$ and $\mathbf{r}_\beta=\mathbf{r}-\mathbf{x}/2$ and interchanging the momentum and radial integrations, the reaction kernel of Equation  \eqref{eq:K2} is rewritten as
\bea
&&\mathcal{K}^{(2)}_{\alpha\beta}(\mathbf{P},\mathbf{q})=\int d^3r e^{i(2\mathbf{P}+\mathbf{q}_{\alpha\beta})\cdot \mathbf{r}}
\times \int d^3x e^{i\mathbf{q}/2\cdot \mathbf{x}}\widetilde{H}_S(\mathbf{r}+\mathbf{x}/2)H_S(\mathbf{r}-\mathbf{x}/2)\\
&& \int \frac{d^3k_\gamma}{(2\pi)^3}g^{(+)}_{\alpha\gamma}(k_\gamma)e^{i\mathbf{k}_\gamma\cdot \mathbf{x}} \nonumber
\eea

The $k_\gamma$-integral results in the coordinate propagator $g_{\alpha\beta}(x)$ defined in Equation  \eqref{eq:gx}. Thus,
\be
\mathcal{K}^{(2)}_{\alpha\beta}(\mathbf{P},\mathbf{q})=\int d^3r e^{i(2\mathbf{P}+\mathbf{q}_{\alpha\beta})\cdot \mathbf{r}}
\int d^3x e^{i\frac{1}{2}\mathbf{q}\cdot \mathbf{x}}\widetilde{H}_S(\mathbf{r}+\mathbf{x}/2)H_S(\mathbf{r}-\mathbf{x}/2)g_{\alpha\beta}(x).
\ee

The form factor product is expanded into multipoles:
\be
\widetilde{H}_S(\mathbf{r}+\mathbf{x}/2)H_S(\mathbf{r}-\mathbf{x}/2)=4\pi\sum_{\ell m}Y_{\ell m}(\hat{\mathbf{r}})Y^*_{\ell m}(\hat{\mathbf{x}})H_{\ell}(r,x)
\ee
with scalar form factors $H_\ell(r,x)$.  Due to symmetry, 
the odd multipoles are strongly suppressed and even vanish identically if the two absorption form factors are equal. Then, only even multipoles $\ell=0,2\ldots$ contribute. In practice, the expansion is dominated by far by the monopole component $H_0(r,x)$. Doing so, the $x$-integral can be performed, leading to
\be
g_0(q|r)=\int^\infty_0 dxx^2 g_{\alpha\beta}(x)j_0(\frac{1}{2}qx)H_0(r,x).
\ee

Since $g_{\alpha\beta}(x)$ is strongly peaked at $xßsim x$, we may replace $H_0(r,x)\approx H_0(r,0)$ as a further approximation, allowing us to extract that form factor from the integral:
\be
g_0(q|r)\approx H_0(r,0)\int^\infty_0 dxx^2 g_{\alpha\beta}(x)j_0(\frac{1}{2}qx)=H_0(r,0)\frac{1}{4\pi} g^{(+)}_{\alpha\beta}(q)
\ee

Finally, as a good approximation, we end up with
\be
\mathcal{K}^{(2)}_{\alpha\beta}(\mathbf{P},\mathbf{q})\approx
 4\pi\int^\infty_0 dr r^2 H_0(q|r,0)j_0(|2\mathbf{P}+\mathbf{q}_{\alpha\beta}|r)\frac{1}{4\pi} g^{(+)}_{\alpha\beta}(q) ,
\ee
where a closer examination of the remaining integral reveals
\bea
4\pi\int^\infty_0 dr r^2 H_0(r,0)j_0(|2\mathbf{P}+\mathbf{q}_{\alpha\beta}|r)&=&\int d^3r e^{i(2\mathbf{P}+\mathbf{q}_{\alpha\beta})\cdot \mathbf{r}}H_0(r,0)\\
&=& 4\pi\int d^3r e^{i(2\mathbf{P}+\mathbf{q}_{\alpha\beta})\cdot \mathbf{r}}\widetilde{H}_S(\mathbf{r})H_S(\mathbf{r}).
\eea

Hence, within the monopole approximation, we find the result:
\be
\mathcal{K}^{(2)}_{\alpha\beta}(\mathbf{P},\mathbf{q})\approx
\int d^3r e^{i(2\mathbf{P}+\mathbf{q}_{\alpha\beta})\cdot \mathbf{r}}\widetilde{H}_S(\mathbf{r})H_S(\mathbf{r})g^{(+)}_{\alpha\beta}(q) ,
\ee
as anticipated in Section \ref{ssec:DisAbs}.

\section[\appendixname~\thesection]{Multipole Expansions of Spin Scalar Operators 
}\label{app:SpinScalar}
The multipoles of the spin scalar operators are
\be
R_0(\mathbf{p}|i)=4\pi\sum_{LM}Y^*_{LM}(\hat{\mathbf{p}})i^LY_{LM}(\hat{\mathbf{r}}_i)j_L(pr_i)
\ee
where $Y_{LM}$ is a rank-L spherical harmonic, and $j_L(x)$ is a spherical Riccati--Bessel function.

Products of the spin scalar operators result in the double-Fermi (FF) operators:
\be
R_{FF}(\mathbf{p},\mathbf{p}'|ij)=(4\pi)^2\sum_{L_1L_2,LM}(-)^{M}\mathcal{Y}_{(L_1L_2)L-M}(\hat{\mathbf{p}},\hat{\mathbf{p}}')
R_{(L_1L_2)LM}(pp'|ij)
\ee
given by the bi-spherical harmonics:
\be
\mathcal{Y}_{(L_1L_2)LM}(\hat{\mathbf{x}},\hat{\mathbf{x}}')=
\sum_{M_1M_2}(L_1M_1L_2M_2|LM)Y_{L_1M_1}(\hat{\mathbf{x}})Y_{L_2M_2}(\hat{\mathbf{x}}')
\ee
and with the spacial multipole operators:
\be
R_{(L_1L_2)LM}(p,p'|ij)=
\mathcal{Y}_{(L_1L_2)LM}(\hat{\mathbf{r}}_i,\hat{\mathbf{r}}_j)j_{L_1}(pr_i)j_{L_2}(p'r_j).
\ee

Thus, the spin--scalar operators will excite the full spectrum of Fermi-like transitions, 
leading to nuclear states of natural parity $J^\pi_A=0^+,1^-,2^+\ldots$, if the operator is acting on a nucleus $A$ with a $0^+$ ground state.

\section[\appendixname~\thesection]{Multipole Expansions of Spin Vector Operators}\label{app:SpinVector}
Spin operators are given, conveniently, on the basis of bi-orthogonal spherical unit vectors $\{\mathbf{e}_\mu,\mathbf{e}^*_\mu\}$:
\be
\mathbf{e}_{\pm 1}=\frac{\pm }{\sqrt{2}}\left(\mathbf{e}_x\pm i\mathbf{e}_y \right)\quad ; \quad \mathbf{e}_0=\mathbf{e}_z ,
\ee
and with the dual unit vectors $\mathbf{e}^*_\mu=(-)^\mu\mathbf{e}_{-\mu}$, we obtain $\mathbf{e}^*_\mu\mathbf{e}_\nu=\mathbf{e}_\mu\mathbf{e}^*_\nu=\delta_{\mu\nu}$. Scalar products of two vector operators $\mathbf{V}_{1,2}=\sum_\mu V^\mu_{1,2}\mathbf{e}^*_\mu$ are given by $\mathbf{V}_1\cdot \mathbf{V}_2=\sum_\mu (-)^\mu V^\mu_1V^{-\mu}_2$, while
$\mathbf{V}^\dag_1\cdot \mathbf{V}_2=\sum_\mu V^{\mu*}_1V^{\mu}_2$.

Using $\bm{\sigma}=\sum_\mu \sigma_\mu \mathbf{e}^*_\mu$,
the spin vector operators are decomposed into spherical components by projection onto $\mathbf{e}_\mu$:
\be
R_\mu(\mathbf{p}|i)=\vec{R}_1(\mathbf{p}|i)\cdot \mathbf{e}_\mu \quad ; \quad
\vec{R}_1(\mathbf{p}|i)=\sum_{\mu=0,\pm 1}R_\mu(\mathbf{p}|i)\mathbf{e}^*_\mu .
\ee
and the spin vector operators become
\be
\vec{R}_1(\mathbf{p}|i)=4\pi\sum_{LM_LJM\mu}Y^*_{LM_L}(\hat{\mathbf{p}})T_{(L1)JM}(p|i)
(LM 1\mu|JM)\mathbf{e}^*_\mu .
\ee

We have introduced the spin orbital multipole operators ($S=0,1$):
\be
T_{(LS)JM}(p|i)=
\left[i^LY_{L}(\hat{\mathbf{r}}_i)\otimes [\bm{\sigma}_i]^S\right]_{JM}j_L(pr_i)
\ee
where
\be
\left[i^LY_{L}(\hat{\mathbf{r}})\otimes [\bm{\sigma}]^S\right]_{JM}=\sum_{M_LM_S}
(LM_LSM_S|JM)i^LY_{LM_L}(\hat{\mathbf{r}})[\sigma_{M_S}]^S .
\ee

Obviously, that definition is, in fact, applicable for both $S=0$ and $S=1$. Applying the triangle rule to $\mathbf{J}=\mathbf{L}+\mathbf{S}$ predicts $|L-S|\leq J \leq L+S$, which for $S=1$ implies $J=|L-1|,L,L+1$. The spin vector operators with $J=|L\pm 1|$ and parity $\pi_J=(-)^{L+1}$ induce transitions to the whole spectrum of unnatural parity nuclear states: $J^\pi_A=0^-,1^+,2^-\ldots$. The spin vector multipole operators for which spin and orbital angular momenta are coupled to $J=L$ have parity $\pi_J=(-)^L$. They contribute to transitions to the nuclear states of natural parity, $J^\pi_A=1^-,2^+\ldots$, by exciting the spin components, which are, in general, part of the wave functions. Dipole transitions are the best known cases and allow us to study the whole spectrum of spin dipole strength distributed over the triplet of $J^\pi_A=0^-,1^-,2^-$ states reached by acting with the spin dipole operator on a nucleus $A$ with a $0^+$ ground state.

The operators encountered in $\Sigma_{GG}$, $\Sigma_{T_nT_n}$ and the mixed $\Sigma_{FT}$ and $\Sigma_{GT}$ are essentially constructed by the above rules, which here were solved 
for scalar products but are easily  generalized to the case of higher-rank tensor products.

The products of spin vector operators appearing in the mixed Fermi--Gamow--Teller FG and FG operators are
\be
R_{GG}(\mathbf{p},\mathbf{p}'|ij)=\vec{R}_1(\mathbf{p},i)\cdot \vec{R}_1(\mathbf{p}',j)
\ee
and require additional steps of recoupling before they are finally obtained as
\bea
R_{GG}(\mathbf{p},\mathbf{p}'|ij)&=&(4\pi)^2\sum_{LM}\sum_{L_1,L_2,J_1,J_2}(-)^{L_1+L_2-J_1}
\widehat{J}_1\widehat{J}_2W(L_1J_1L_2J_2;1L)\nonumber\\
&\times&\mathcal{Y}^*_{(L_1L_2)LM}(\hat{\mathbf{p}},\hat{\mathbf{p}}')R^{(L_1L_2,J_1J_2)}_{LM}(pp'|ij)
\eea
where $\mathcal{Y}^*_{(L_1L_2)LM}=(-)^M\mathcal{Y}_{(L_1L_2)L-M}$. Reduced spin orbital multipole operators are introduced:
\bea
R^{(L_1L_2,J_1J_2)}_{LM}(pp'|ij)&=&
\left[T_{(L_11)J_1}(p|i)\otimes T_{(L_21)J_2}(p'|j)\right]_{LM}\nonumber\\
&=&\sum_{M_1M_2}(J_1M_1J_2M_2|LM)
T_{(L_1S)J_1M_1}(p|i)T_{(L_2S)J_2M_2}(p'|j)
\eea
where $W(L_1J_1L_2J_2;1L)$ is a Racah coefficient.

The mixed products of spin scalar and spin vector operators acting as two-body operators in the same nucleus are found by the same recoupling techniques. Using $\bm{\sigma}=\sum_\mu\sigma_\mu \mathbf{e}*_\mu$, we obtain
\bea
&&\vec{R}_{FG,ab}(\mathbf{p},\mathbf{p}')=(4\pi)^2\sum_{\mu}\mathbf{e}*_\mu
\sum_{L_1L_2,LM_L}\mathcal{Y}^*_{(L_1L_2)LM_L}(\hat{\mathbf{p}},\hat{\mathbf{p}}')\nonumber\\
&&\times (-)^{L_1+L_2-L}(L_1M_1L_2M_2|LM_L)(L_21\mu|J_2N_2)(L_1M_1J_2N_2|JM)R^{(L_1L_2J_2)}_{JM,ab}(pp').
\eea

The reduced multipoles are
\be
R^{(L_1L_2J_2)}_{JM,ab}(pp')=\sum_{(13)}\left[T_{(L_10)L_1}(p|1)\otimes T_{(L_21)J_2}(p'|3) \right]_{JM}.
\ee

With $(L_2M_21\mu|J_2N_2)=\frac{\widehat{J}_2}{\sqrt{3}}(-)^{-J_2-1+N_2+\mu}(L_2M_2J_2-N_2|1-\mu)$, the summation over $N_2$ can be performed:
\bea
&&\sum_{N_2}(-)^{N_2}(L_1M_1J_2N_2|JM)(L_2J_2-N_2|1-\mu)(L_1M_1L_2M_2|LM_L)\nonumber\\
&&=(-)^{L_2+J-L}\widehat{J}\sqrt{3}W(L_1JL_21;J_2L)(JM1-\mu|LM_L)
\eea
and $(JM1-\mu|LM_L)=(-)^{-1+M_L+L-\mu}\frac{\widehat{L}}{\sqrt{3}}(JML-M_L|1\mu)$. Collecting phases and pre-factors, the final result is:
\bea
&&\vec{R}_{FG,ab}(\mathbf{p},\mathbf{p}')=(4\pi)^2
\sum_{L_1L_2J_2,LM_L,JM}\mathcal{Y}^*_{(L_1L_2)LM_L}(\hat{\mathbf{p}},\hat{\mathbf{p}}')\\
&&\times (-)^{L_1-L+J-J_2+M_L}\frac{\widehat{J}\widehat{J}_2\widehat{L}}{\sqrt{3}}W(L_1JL_21;J_2L)R^{(L_1L_2J_2)}_{JM,ab}(pp')\nonumber
\sum_{\mu}(JML-M_L|1\mu)\mathbf{e}_\mu .
\eea

The elements $\vec{R}_{GF,ab}$, $\vec{R}_{FG,AB}$ and $\vec{R}_{FG,BA}$ are obtained by the same approach.

\section[\appendixname~\thesection]{Recoupling of Bi-Spherical Harmonics}\label{app:BiSpherical}

An important property of bi-spherical harmonics is that for $\hat{\mathbf{x}}=\hat{\mathbf{x}}'$, they reduce
to ordinary spherical harmonics:
\be
\mathcal{Y}_{(L_1L_2)LM}(\hat{\mathbf{x}},\hat{\mathbf{x}})=A_{L_1L_2L}Y_{LM}(\hat{\mathbf{x}})
\ee
with
\be\label{eq:A3L}
A_{L_1L_2L}=\frac{\widehat{L}_1\widehat{L}_2}{\sqrt{4\pi}\widehat{L}}(L_10L_20|L0)
\ee

The Clebsch--Gordan coefficient vanishes if $L_1+L_2+L=2n+1$ is an odd number.

Evaluation of the DSCE-NME leads to products of two bi-spherical harmonics in the momenta $\mathbf{p}$ and $\mathbf{p}'$. By further steps of recoupling, the resulting product of four ordinary spherical harmonics can be reduced to a product of two spherical harmonics of the same argument, finally forming a single bi-spherical harmonic.
\bea
&&\mathcal{Y}_{(L_1L_2)LM}(\hat{\mathbf{p}},\hat{\mathbf{p}}')\mathcal{Y}_{(L_3L_4)L'M'}(\hat{\mathbf{p}},\hat{\mathbf{p}}')=
\\
&&\sum_{\ell\ell'\lambda\mu}A_{L_1L_3\ell}A_{L_2L_4\ell'}
\mathcal{Y}_{\ell\ell'\lambda\mu}(\hat{\mathbf{p}},\hat{\mathbf{p}}')\sum_{\lambda'\mu'}(LML'M'|\lambda'\mu')
\sum_{M_1M_2M_3M_4mm'}\nonumber\\
&&\times(L_1M_1L_2M_2|LM)(L_3M_3L_4M_4|L'M')(L_1M_1L_3M_3|\ell m) \nonumber\\
&&\times(L_2M_2L_4M_4|\ell' m')(LML'M'|\lambda'\mu')(\ell m \ell' m'|\lambda \mu)\nonumber
\eea
which results in 9-j symbol:
\bea
&&\mathcal{Y}_{(L_1L_2)LM}(\hat{\mathbf{p}},\hat{\mathbf{p}}')\mathcal{Y}_{(L_3L_4)L'M'}(\hat{\mathbf{p}},\hat{\mathbf{p}}')=
\\
&&\sum_{\ell\ell'\lambda\mu}
\mathcal{Y}_{(\ell\ell')\lambda\mu}(\hat{\mathbf{p}},\hat{\mathbf{p}}')(LML'M'|\lambda\mu)
A_{L_1L_3\ell}A_{L_2L_4\ell'}\widehat{\ell}\widehat{\ell}'\widehat{L}\widehat{L}'
\left\{
  \begin{array}{ccc}
    L_1 & L_2 & L \\
    L_3 & L_4 & L' \\
    \ell & \ell' & \lambda \\
  \end{array}
\right\}\nonumber
\eea

Defining
\be
\Gamma^{L_1L_2L}_{L_3L_4L'}(\ell\ell'\lambda)=\frac{1}{4\pi}
\widehat{L}_1\widehat{L}_2\widehat{L}_3\widehat{L}_4\widehat{L}\widehat{L}'
\left\{
  \begin{array}{ccc}
    L_1 & L_2 & L \\
    L_3 & L_4 & L' \\
    \ell & \ell' & \lambda \\
  \end{array}
\right\}\nonumber
\ee
we obtain
\bea
\mathcal{Y}_{(L_1L_2)LM}(\hat{\mathbf{p}},\hat{\mathbf{p}}')\mathcal{Y}_{(L_3L_4)L'M'}(\hat{\mathbf{p}},\hat{\mathbf{p}}')=
\sum_{\ell\ell'\lambda\mu}\Gamma^{L_1L_2L}_{L_3L_4L'}(\ell\ell'\lambda)(LML'M'|\lambda\mu)
\mathcal{Y}_{(\ell\ell')\lambda\mu}(\hat{\mathbf{p}},\hat{\mathbf{p}}')
\eea
and find
\bea
\left[\mathcal{Y}_{(L_1L_2)L}(\hat{\mathbf{p}},\hat{\mathbf{p}}')\otimes\mathcal{Y}_{(L_3L_4)L'}(\hat{\mathbf{p}},\hat{\mathbf{p}}')\right]_{\lambda\mu}=
\sum_{\ell\ell'}\Gamma^{L_1L_2L}_{L_3L_4L'}(\ell\ell'\lambda)
\mathcal{Y}_{(\ell\ell')\lambda\mu}(\hat{\mathbf{p}},\hat{\mathbf{p}}').
\eea

The bi-spherical harmonics form an over-complete system, but there is an orthogonality relation:
\bea
\int d\hat{\mathbf{p}}\int d\hat{\mathbf{p}}'
\mathcal{Y}^*_{(L_1L_2)\lambda\mu}(\hat{\mathbf{p}},\hat{\mathbf{p}}')
\mathcal{Y}_{(L_3L_4)\lambda'\mu'}(\hat{\mathbf{p}},\hat{\mathbf{p}}')&=&\\
\sum_{M_1M_2}(L_1M_1L_2M_2|\lambda\mu)(L_1M_1L_2M_2|\lambda'\mu')&=&\delta_{\lambda\lambda'}\delta_{\mu\mu'}\nonumber .
\eea


\begin{adjustwidth}{-\extralength}{0cm}

\printendnotes[custom]
\reftitle{References} 

\PublishersNote{}
\end{adjustwidth}

%
\end{document}